\documentclass[journal]{IEEEtran}

\ifCLASSINFOpdf
\usepackage[pdftex]{graphicx}
\graphicspath{{./graph/}}
\DeclareGraphicsExtensions{.pdf,.jpeg,.png}
\else
\usepackage[dvips]{graphicx}
\graphicspath{{./graph/}}
\DeclareGraphicsExtensions{.eps}
\fi

\usepackage{dblfloatfix}
\usepackage{graphicx}
\usepackage{epstopdf}
\usepackage{algorithm}
\usepackage[vlined, ruled, algo2e]{algorithm2e}
\usepackage{booktabs}
\usepackage{multirow}
\usepackage[table,xcdraw]{xcolor}
\usepackage{mdwmath}
\usepackage{amssymb}
\usepackage{amsmath}
\usepackage{pbox}
\usepackage{csquotes}
\usepackage{hyperref}
\usepackage{cite}

\newcommand*\rot{\rotatebox{90}}

\begin{document}
\title{Sensor Fusion for Public Space Utilization Monitoring in a Smart City}
\author{Billy~Pik~Lik~Lau,
        Nipun~Wijerathne,
        Benny~Kai~Kiat~Ng,
        and~Chau~Yuen,
\thanks{Billy Pik Lik Lau (Corresponding Author), Nipun Wijerathne, Benny Kai Kiat Ng, and Chau Yuen are with the Engineering Product and Design, Singapore University of Technology and Design, e-mail:
	 billy\_lau@mymail.sutd.edu.sg, yuenchau@sutd.edu.sg.}
\thanks{}}

\markboth{}%
{Shell \MakeLowercase{\textit{et al.}}: Bare Demo of IEEEtran.cls for IEEE Journals}
\maketitle


\begin{abstract}
Public space utilization is crucial for urban developers to understand how efficient a place is being occupied in order to improve existing or future infrastructures. 
In a smart cities approach, implementing public space monitoring with Internet-of-Things (IoT) sensors appear to be a viable solution.
However, choice of sensors often is a challenging problem and often linked with scalability, coverage, energy consumption, accuracy, and privacy.
To get the most from low cost sensor with aforementioned design in mind, we proposed data processing modules for capturing public space utilization with Renewable Wireless Sensor Network (RWSN) platform using pyroelectric infrared (PIR) and analog sound sensor.
We first proposed a calibration process to remove false alarm of PIR sensor due to the impact of weather and environment. 
We then demonstrate how the sounds sensor can be processed to provide various insight of a public space.
Lastly, we fused both sensors and study a particular public space utilization based on one month data to unveil its usage.

\end{abstract} 
\begin{IEEEkeywords}
Smart City, Internet of Things, Sensor Fusion, Space Occupancy, Public Space Utilization, Spatial-Temporal.
\end{IEEEkeywords}

\section{Introduction}
\label{sec:introduction}

\IEEEPARstart{S}{mart} city vision provides authorities to manage the city's asset intelligently and help the city planner better design the city to serve the residents.
The advancement of Internet-of-Things (IoT) and big data enable various information to be collected remotely and send to cloud for further processing. 
Interconnectivity between different devices is crucial in smart city and extensive standards with deployment practices for data transmission has been studied in \cite{zanella2014internet}.
Although standardizing interconnected devices protocol still remain a challenge, most issues renders down to the application and insights generated through the implementation of the IoT devices. 

Therefore, wireless sensor networks (WSN) have been developed to address the large scale deployment in almost any environment ranging from factory to urban city. While reducing any physical cable communication, it also allows different sensor nodes to be installed at any corner to perform any designated tasks. It is well suited for smart city application due to its scalability, robustness, and coverage.
Example of  WSN application in urban area are listed as follows: City Wide Deployment \cite{zanella2014internet,lee2008intelligent,campbell2008rise}, 
Smart Metering \cite{fadel2015survey,chi2016zigbee}, Security Surveillance System \cite{eigenraam2016smart}, health-care \cite{aziz2016smart}, and etc.

The motivation of this paper is to study the utilization of an urban public space using IoT concept. 
By introducing sensor nodes to collected data, automation process benefits urban developer to better understand public space.
It reduces labor intensive on-site investigation and provide an unbiased quantitative measurement on the open space occupancy study.
As the crowd density lived in urban area is getting higher, urban planning is critical in order to make full use of every land estates especially in a small country such as Singapore. 
Thus, it is important that urban developers understand the usage existing infrastructure in order to better serve the residents. 
Previously in our works \cite{lau2016spatial}, we show that space utilization can be used to generate certain degree of insights. 
These insights can be used to improve urban planning to create a more live-able place for urban residents. 

Space utilization is also commonly known as space occupancy and various sensors have been studied to perform such tasks.
Recent space utilization topic revolves around heat sensing such as thermal imaging or temperature grid \cite{tyndall2016occupancy} to estimate the space occupancy. 
They have studied different type of classifiers to estimate the space occupancies based on temperature variation. 
In \cite{raykov2016predicting}, Yordan et al. studied the pyroelectric infrared (PIR) sensor and used it for room occupancies detection.
On the other hand, sounds can be used for detecting activity or events have been studied in \cite{salamon2015unsupervised, heittola2013context} and can be used in conjunction to study space utilization. 
Moreover, density based space occupancy such as sensing crowd density at a specific location have been conducted in \cite{li2015senseflow,schauer2014estimating} but they are more people centric.
Computer vision approaches such as camera proposed in \cite{munaro2012tracking,spinello2011people,junior2010crowd,chan2008privacy} also can be used in study a pinpoint area of utilization and other information such as crowd density. 
However, it is computational expensive, which requires better hardware and may lead to cost issue in large scale deployment. 
Also, power needed for processing image from camera is relatively high and the solar panel installed on light pole is not able to meet the power consumption requirement. 
Alternate method for off-line processing would be sending back data to cloud but that would require massive storage and communication bandwidth, where cost of implementation is high. Main concern of the computer vision techniques are mostly intrusive, where it is challenging to seek necessary clearance for its installation at public area due to privacy concern. 
Finally, vision based solution may have difficulty at low light condition during night time.
Based on the survey conducted earlier, we chose sounds and PIR sensor as the main sensor for detecting activity in public space.

\begin{figure*}[ht]
	\includegraphics[trim = 0 11.5cm 0 0,width=0.98\textwidth]{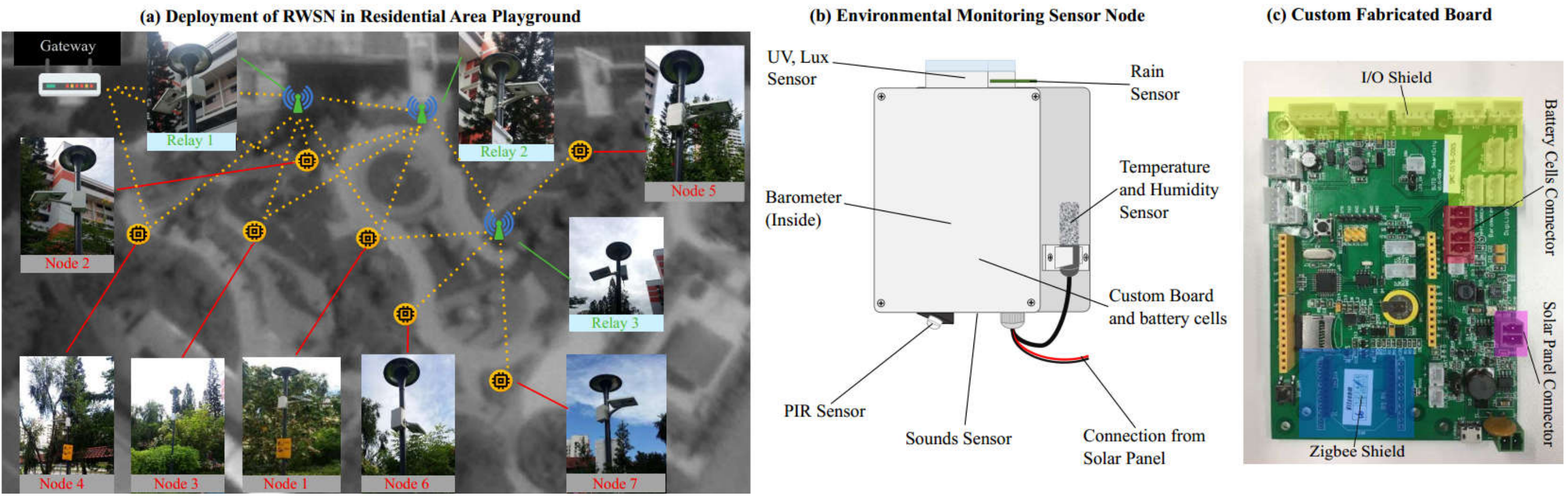}
	\vspace{-0.5em}	
	\centering \caption{Overall hardware architecture of the RWSN (From left to right): (a) an example of place where the sensor nodes are being deployed, (b) the anatomy of the sensor nodes, and (c) Custom fabricated board}
	\vspace{-1.0em}	
	\label{fig:MixFigureHardware}
\end{figure*}

In this paper, we proposed data processing model to process motion and noise data collected from Renewable Wireless Sensor Networks (RWSN). 
We have deployed RWSN in residential area and data has been collected to study the public space utilization.
We address some of the challenges of computing public space utilization such as data calibration, rain, and activity detection using sounds sensor.
Data calibration is proposed to eliminate the false positive generated by the motion sensor, as the PIR sensor is sensitive to light and high temperature. 
The analog sounds sensor collecting histogram of noise level in more granularity details in order to detect various activity happening in a point-of-interest (PoI). 
After defining both motion and sounds sensor, we fused both motion and sounds sensor to provide a broader range of public space utilization.
Lastly, we applied the space utilization computation models using data over 1 month (August 2016) to study public space utilization on playground area.

Our contributions in this paper can be summarized as following:
\vspace{-1.4mm}
\begin{itemize}	
	\item We managed to correctly detect public space utilization by motion sensor using the data calibration module to remove any false positive by correlating with the key factors.
	\item We use sounds sensor to detect public space utilization with signal processing techniques such as Principal Component Analysis (PCA) and wavelet transform. Also, we noticed our noise sensors are capable to distinguish between raining and normal day.
	\item Utilizing the ideology of sensor fusion, we combine motion and sounds sensor to compute a more complete public space utilization.
\end{itemize}
\vspace*{-1.4mm}

The rest of the paper is organized as follows:
In section \ref{sec:ArchitectureaAndData_Processing}, we briefly introduce RWSN hardware components and discuss about overall data processing model. 
In Section \ref{sec:MotionUtz}, we present the calibration model for the motion sensor to adapt the changes made to the sensor node. 
Subsequently, process of computing space utilization using sounds sensor is detailed in Section \ref{sec:utilizationSound}. Sensor fusion between sounds and motion sensor is presented in Section \ref{sec:combineUtz}. 
Lastly, we conclude our works in Section \ref{sec:conclusion}.

\vspace*{-2.0mm}

\section{System Overview}
\label{sec:ArchitectureaAndData_Processing}
\subsection{Renewable Wireless Sensor Network}
\label{subsec:RWSN_Hardware}
The proposed RWSN is an extension of WSN equipped with renewable energy instead of relying on either battery or power outlet. 
It is deployed in public space of a residential area in Singapore to study the space utilization, an example of PoI, which is a playground as is shown in Fig. \ref{fig:MixFigureHardware}(a). 
Three relays and seven sensor nodes are installed at the light pole around the premise of playground. 
Gateway is installed at the switch room at residential area nearby to transmit the data collected to cloud. 

Fig. \ref{fig:MixFigureHardware}(b) and \ref{fig:MixFigureHardware}(c) shows the anatomy of the sensor nodes and custom fabricated board respectively. 
Several notable features of the custom fabricated board includes battery cells and solar panel installation, which allow it to be self sustained during data collection. 
The solar panel with maximum efficiency of 10w$\pm5\%$ around are able to charge four NCR—18650B batteries through out the broad day light in a tropical country such as Singapore. 
Further investigation of power consumption, network traffic, scalability, and stability of the RWSN sounds interesting and will be included in our future works.
Next, we are going to discuss three different hardwares that form the very core of RWSN, which are (1) Gateway, (2) Relay, and (3) Environmental Monitoring Sensor Node.

\subsubsection{Gateway (External Power Source)}
The gateway will be acting as an intermediate between database and all the sensor nodes. 
Gateway is composed of Xbee receiver and Raspberry Pi Microprocesser, which will process the data packet sent by sensor nodes and upload to the database accordingly.
Mesh network centering the gateway will be formed in order to provide a larger coverage for the sensor nodes.

\subsubsection{Relay (Energy Harvesting)}
Transmitted signal may have limited range due to the interference from structural design and other radio signal. 
In order to mitigate the signal degradation, a Xbee relay is used as a repeater to increase the coverage of the wireless mesh network.
The hardware component of the relay composed of batteries, charging module, Xbee antenna, and solar panel. 

\subsubsection{Environmental Monitoring Sensor Node (Energy Harvesting)}
The environmental monitoring sensor node has the capabilities to monitor surrounding information using an Arduino compatible board and Xbee transmitter. 
The environmental information recorded are following: barometer, temperature, light intensity, resistive rain, ultra-violet (UV) index, humidity, motion, and noise sensor. 
The rain, light intensity, and UV sensor are placed on the top of the sensor box to generate a more accurate reading.
PIR and noise sensor are placed at the bottom of the sensor box pointing to ground. 
Note that, motion and sound sensors are actively accumulating data reading for 5 minutes interval, where as other sensors only consider snapshot of data every 5 minutes.
External solar panel is installed and able to supply power to the sensor box and charge the battery simultaneously.

\begin{table}[t]
	\centering
	\caption{Sensor Data Specification and Sampling Rate}	
	\fontsize{7pt}{7pt}\selectfont
	\setlength{\textfloatsep}{10pt plus 1.0pt minus 2.0pt}	
	\label{tbl:sensorSpec}
	\begin{tabular}{@{}lll@{}}
		\toprule
		Sensor & Data Range & Sampling Rate \\ \midrule
		Motion, $M_t$, & 0 - 1 (Relative Value) & \begin{tabular}[c]{@{}l@{}}Accumulative every 3\\ seconds for 5 minutes\end{tabular} \\
		Noise, $X_t$ & \begin{tabular}[c]{@{}l@{}}Histogram for 5 different bins \\  
			$\{0\text{-}6,6\text{-}10,10\text{-}20,20\text{-}50,>50\}$ \end{tabular} & \begin{tabular}[c]{@{}l@{}}Accumulative every 0.1\\ seconds for 5 minutes\end{tabular} \\
		Temperature, $K_t$ & -10 - 80$^{\circ}$c & \begin{tabular}[c]{@{}l@{}}Snapshot every 5 minutes\end{tabular} \\
		Lux, $L_t$& 0.1 - 40000 Lux & \begin{tabular}[c]{@{}l@{}}Snapshot every 5 minutes\end{tabular} \\
		Rain, $R_t$& \begin{tabular}[c]{@{}l@{}}0 - 1024\end{tabular} & \begin{tabular}[c]{@{}l@{}}Snapshot every 5 minutes\end{tabular} \\
		UV, $U_t$& \begin{tabular}[c]{@{}l@{}}0 - 1024\end{tabular} & \begin{tabular}[c]{@{}l@{}}Snapshot every 5 minutes\end{tabular} \\
		Barometer, $B_t$& 300 - 1200mbar & \begin{tabular}[c]{@{}l@{}}Snapshot every 5 minutes\end{tabular} \\
		Humidity, $H_t$& 0 - 100\% RH & \begin{tabular}[c]{@{}l@{}}Snapshot every 5 minutes\end{tabular} \\
		\bottomrule
	\end{tabular}
\end{table}

\vspace*{-2.4mm}

\subsection{Data Processing}

\begin{figure}[b]
	\vspace{-0.5em}	
	\includegraphics[width=0.48\textwidth]{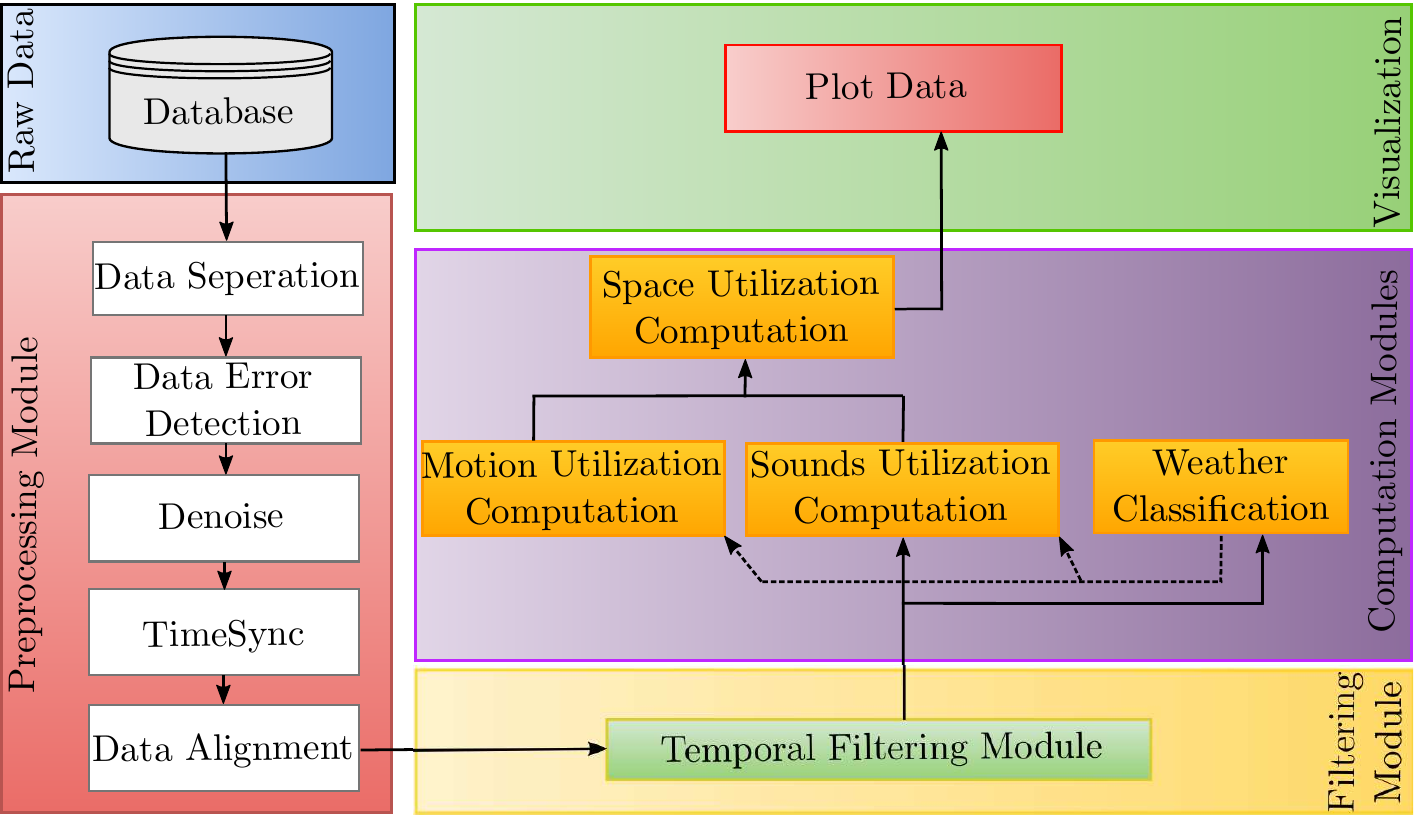}
	\vspace{-1em}
	\centering \caption{The data processing modules for the RWSN}
	\vspace{-1.0em}
	\label{fig:RWSN_Architcture}
\end{figure}

In this paper, we will highlight about data processing modules for computing the space utilization using PIR and sound sensor.
Detail of data collected is presented in Table \ref{tbl:sensorSpec} with the respective data value range and frequency. 
Next, We characterize the data collected as a data set consist of $S_{t}=\{R_{t},K_{t},H_{t},B_{t},L_{t},U_{t}\}$ as weather perception sensor, while $U_t=\{M_t,X_t\}$ as sensors for monitoring utilization.

Data processing process can be divided into five different components namely, (1) data storage extraction, (2) preprocessing, (3) data filtering, (4) data computation, and (5) data visualization.

First, raw data are downloaded from cloud and separated based on different node unique identity (UID) into matrix for easier processing using data separation module. 
Subsequently, we have introduce \textit{TimeSync} modules to remove duplicated data as well as synchronizing the sensor data by fitting them into the predefined time windows. 
To eliminate the duplicated data received by database, we applied \textit{Denoise} and remove it before feeding it into computation and filtering module. 
The details of the aforementioned modules (Denoise and TimeSync) can be found in \cite{lau2016spatial}.

In this phase, we introduce \textit{Data Error Detection} to eliminate defective data uploaded by gateway due to damaged packet received. 
Each node has specific identifier in order to indicate the sensor ID and values, which are arranged in a specific format of $\{\text{ID:Value}\}$. 
For instance, the gateway may received garbage data from the sensor (either due to retransmission of long packet over Xbee network due to multi-hops relaying). 
This phenomenon causes the downloaded raw data contains invalid syntax such as sensor id and values, which does not match the preassigned list of sensor id and data specification. 
Thus, we will discard the garbage data portion in order to cleanse the data before other preprocessing modules.

Data alignment will arrange the data into two dimensional matrix, which the row indicates time series data and column indicates different types of sensor input. 
This allow us to study different kind of sensor reading at the same temporal domain since all sensors are synchronized using TimeSync. 
The output of the module will be used for filtering and computation module.

During filtering module phase, temporal filtering is applied to select particular temporal space utilization. 
For instance, user is able to choose from ``particular day", ``weekday", ``weekend", or ``no filter" and to feed the filtered data into computation module.

The data computation modules are composed of multiple different sub modules such as motion/sound utilization, combined space utilization, and weather classification. 
Weather Classification module is used to classify the localized weather state of the sensor and is previously discussed in our work at \cite{lau2016spatial}. 
Depending on the user choice of weather to study, motion and sounds utilization module will be sorted based on weather state in order to compute public space utilization individually. 
After that, three different type of public space utilization can be computed, which are motion, sounds, and combined. 
The last module is visualization module, where the public space utilization is displayed in a more graphical way to generate insights.

\section{Space Utilization by Motion Sensor}
\label{sec:MotionUtz}
In this section, we are going to focus public space utilization computation module.  
The motion data captured by the sensor node is based on PIR sensor and it actively detect false alarm during broad daylight. 
To this end, we proposed calibration method to mitigate the false alarm and detect activity correctly. 
We evaluate the motion sensor calibration method and shows it efficiency of removing false alarm without removing the capability of detecting activity.
\vspace*{-3.0mm}

\subsection{Motion Data Calibration}
\label{subsec:dataProcessing}
Due to the design of IR sensor's perception lens, there is a lot false alarm during daytime especially during afternoon. 
It is studied in \cite{hong2013reduction}, it is possible to remove the false alarm given that adaptive threshold is provided. 
To this end, we propose a calibration module to remove the false alarm and correctly identifying activity incurred by nearby residents.

\begin{table}[ht]
	\centering
	\caption{Pearson Correlation between Different Sensors and Motion}
	\label{tbl:PearsonCorrelationSensorMotion}
	\begin{tabular}{@{}l|llll@{}}
		\toprule
		Sensor Reading & \begin{tabular}[c]{@{}l@{}}Node 1\\ Motion\end{tabular} & \begin{tabular}[c]{@{}l@{}}Node 2\\ Motion\end{tabular} & \begin{tabular}[c]{@{}l@{}}Node 4\\ Motion\end{tabular} & \begin{tabular}[c]{@{}l@{}}Node 5\\ Motion\end{tabular} \\ \midrule
		Light, $L_t$ & {\color[HTML]{FE0000} 0.8541} & {\color[HTML]{FE0000} 0.7273} & {\color[HTML]{FE0000} 0.6553} & {\color[HTML]{FE0000} 0.7751} \\
		Temperature, $K_t$ & {\color[HTML]{FE0000} 0.8019} & {\color[HTML]{FE0000} 0.7351} & {\color[HTML]{FE0000} 0.7498} & {\color[HTML]{FE0000} 0.765} \\
		Humidity, $H_t$ & {\color[HTML]{FE0000} -0.8061} & {\color[HTML]{FE0000} -0.7795} & {\color[HTML]{FE0000} -0.7090} & {\color[HTML]{FE0000} -0.7636} \\
		UV, $U_t$ & 0.5593 & 0.5561 & 0.4086 & 0.4381 \\
		rain, $R_t$ & 0.1628 & 0.1413 & 0.1176 & 0.0915 \\
		barometer, $B_t$ & -0.0594 & -0.0328 & 0.2305 & 0.0313 \\ \bottomrule
	\end{tabular}
\end{table}

First, we study the Pearson correlation relationship between the motion sensor and the other environmental sensor as shown in the Table \ref{tbl:PearsonCorrelationSensorMotion}. 
It can be observed that the main factors affecting the motion to fluctuated is highly correlated with the light, temperature, and humidity. 
On site investigation shows that false positive will occur during bright day light, while it is unlikely to happen at night.

Using Nodes 1, 2, 4, and 5 as example, we have identified three sensors that highly correlates with the motion sensor, which are (1) light sensor, (2) temperature, and (3) humidity. 
Observation over 20 days since 10 July until 01 August 2016 are used to study the correlation relationship between the sensor data readings.
Since the humidity is negatively correlates with temperature, we only consider light sensor and temperature as input parameter for the calibration module to reduce the complexity.

\begin{figure}[ht]
	\centering
	\vspace{-0.5em}
	\includegraphics[width=0.4\textwidth]{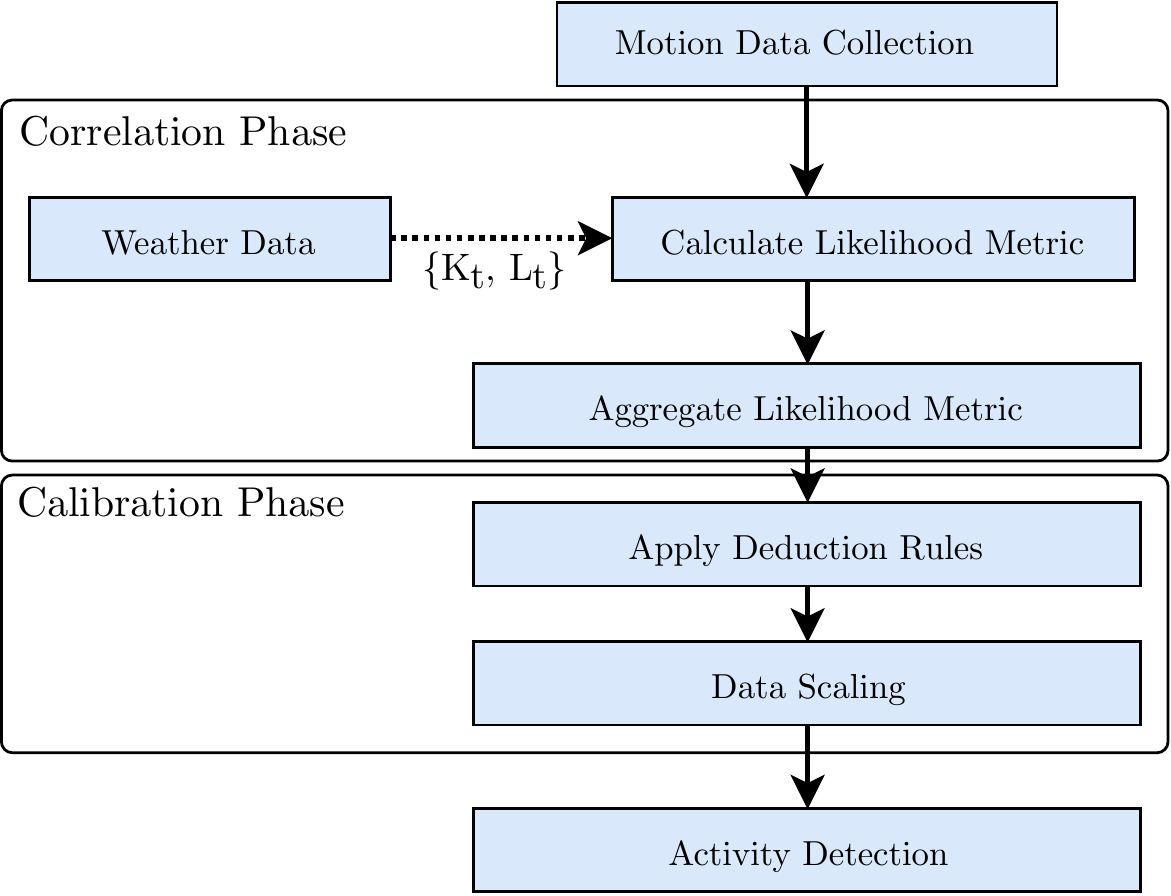}
	\vspace{-1.0em}	
	\caption{Motion Data Processing Model}
	\label{fig:motionDataProcessingModel}
\end{figure}

Since we have identified the critical factors that false alarm is likely to happen and hence we will eliminate the false alarm by deducting off the extra motion value from the PIR sensor. 
We have proposed the motion data processing phase as shown in Fig. \ref{fig:motionDataProcessingModel}, which consists of two phases, namely correlation phase and calibration phase. The correlation phase is to calculate the likelihood that false alarm is likely to happen. 
The data calibration uses concept of deduction to remove false alarm by reducing value based on the likelihood function.

In the correlation phase, the likelihood functions compute how the current weather situation is similar with false alarm weather and determine the deduction value to remove false alarm. 
First, we define the likelihood function for computing sensor values as follow:
\vspace*{-1.0mm}
\begin{equation}
f({S_t}) = \left\{ \begin{matrix}
1.0 & {\text{if}} & {S_t} \ge {\delta _{high}} \hfill \cr
{{{S_t} - {\delta _{low}}} \over {{\delta _{high}} - {\delta _{low}}}} & {\text{if}} & {\delta _{low}} < {S_t} < {\delta _{high}} \hfill \cr 
0.0 & {\text{if}} & {S_t} \le {\delta _{low}} \end{matrix}  \right. ,
\label{eqn:LikelihoodFunction}
\end{equation}
\vspace*{-1.0mm}
where $f(S_{t})$ represents likelihood metric that is based on the upper-bound and lower-bound of the sensor reading. 
We chose temperature $K_t$ and lux sensor $L_t$ as the sensor to determine the likelihood input value because of their highly correlated relationship with motion sensor.
Temperature range of 28$^{\circ}$c to 40$^{\circ}$c and Lux index of 8000 to 33000 are used as input of the likelihood function.
The motion sensor value reading from the node will determine likelihood the false alarm is being recreated and this is based on linear equation. 
The higher value obtain from likelihood function indicate higher chance of false alarm is being occurred.
By obtaining each sensor likelihood value, we can formulate the aggregated likelihood function by as follows:
\vspace*{-1.0mm}
\begin{equation}
p(\alpha ) = \sum\limits_{i = 1}^n {{{f({S_i})} \over n}} {\text{  where }}S_{i} = \{ {L_t},{K_t}\} ,
\label{eqn:sensorPsum}
\end{equation}
where $f(S_i)$ represents the aggregated function where it aggregate the likelihood value from the input sensors. It will be averaged based on the number of sensor input, $n$. 
The final value will be used to decide deduction value for the calibration module. 

\begin{figure}[ht]
	\centering
	\vspace{-0.5em}
	\includegraphics[width=0.41\textwidth]{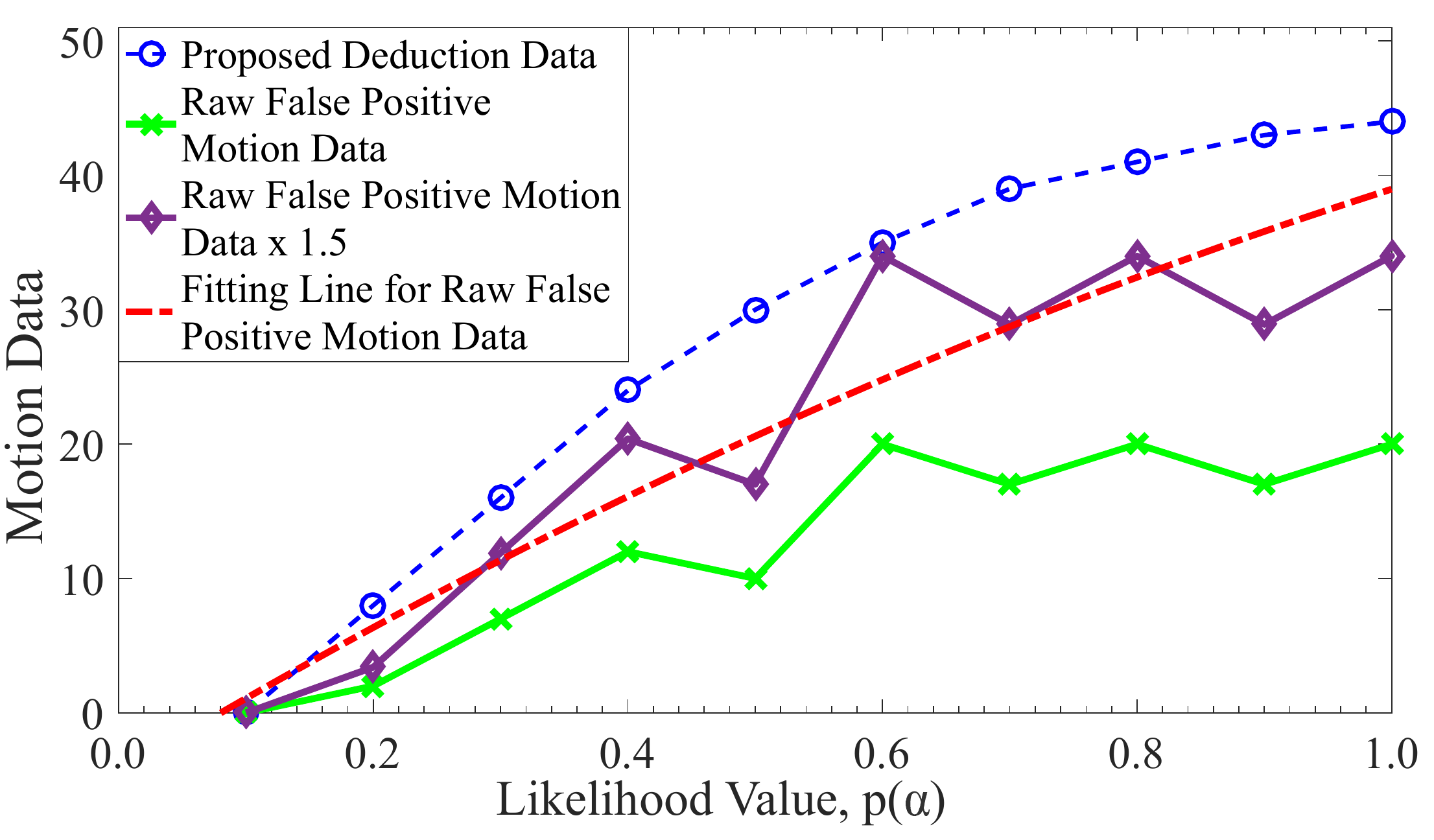}	
	\vspace{-1.0em}		
	\caption{Study of Deduction Value based on Averaged Motion Data}
	\label{fig:DeductionValues}
\end{figure}

The deduction value $D_t$ is decided based on the following equation:
\vspace*{-1.0mm}
\begin{equation}
D_{t} = \left\{ \begin{matrix}
d_{p(\alpha )} & \text{if}&\lambda _{low_{i}} < p(\alpha ) < \lambda _{high_{i}}  \\ 
0 & \text{if}&  p(\alpha) \le \lambda _{low}\\		
\end{matrix}  \right. ,
\label{eqn:deductionRules}
\end{equation}

where $d_{p(\alpha)}$ is the deduction value based on the lower-bound $\lambda _{low}$ and upper-bound $\lambda _{high}$ of the false alarm data collected with ground truth. 
The upper bound and lower bound are decided by the categorical data $d_{t}$ based on the raw data that are marked as false alarm. The raw values used for Eqn \ref{eqn:deductionRules} presented at Fig. \ref{fig:DeductionValues}, where there are 10 different upper-bound and lower-bound can be obtained.

\vspace*{-2mm}

Based on the raw data collected, we set a higher fitting line for the deduction value as the false alarm value tends to be on the higher value. 
It is observed that the raw motion data is not projected in linear function and series of experiment has been conducted shows 50\% more deduction value yields a better result. 
Hence, we applied curve fitting line (polynomial third degree) and smooth out the increased deduction value. 
The proposed deduction value used in the calibration model is displayed in blue color.

Subsequently, we will use the deduction value to calibrate motion sensor using the following equation.
\vspace*{-1.0mm}
\begin{equation}
{m'_t} = \max \left( {{{m}_t} - {D_t}},0 \right),
\label{eqn:deductionFunction}
\end{equation}
where ${m}_t$ represents the raw motion value and ${m'_t}$ is the calibrated motion value. 
It has a lower-bound of 0, where it will always guarantee the calibrated motion value is a positive real number.

Since we have deducted some value from the raw motion data, it would not be a legit comparison with other non-deducted value. 
To compensate motion value removed from the calibration module, we will apply the scaling function to the calibrated motion data. Hence, scaling equation at below is applied to ${m'_t}$ to generated processed motion data, $\mathcal{M}_{t}$.
\vspace*{-1.0mm}
\begin{equation}
\mathcal{M}_{t} = {m'_t}\times\left( {{{G} \over {G - {D_t}}}} \right),
\label{eqn:scalingMotion}
\end{equation}
where $G$ is the maximum motion value can be recorded over 5 minutes sampling rate. Using the specification of the PIR sensor, maximum input of the motion value would be 100. We have defined $G = 100$ and scale back de calibrated motion data.
\vspace*{-4.0mm}

\begin{figure}[ht]
	\centering
	\vspace{-0.5em}	
	\includegraphics[width=0.42\textwidth]{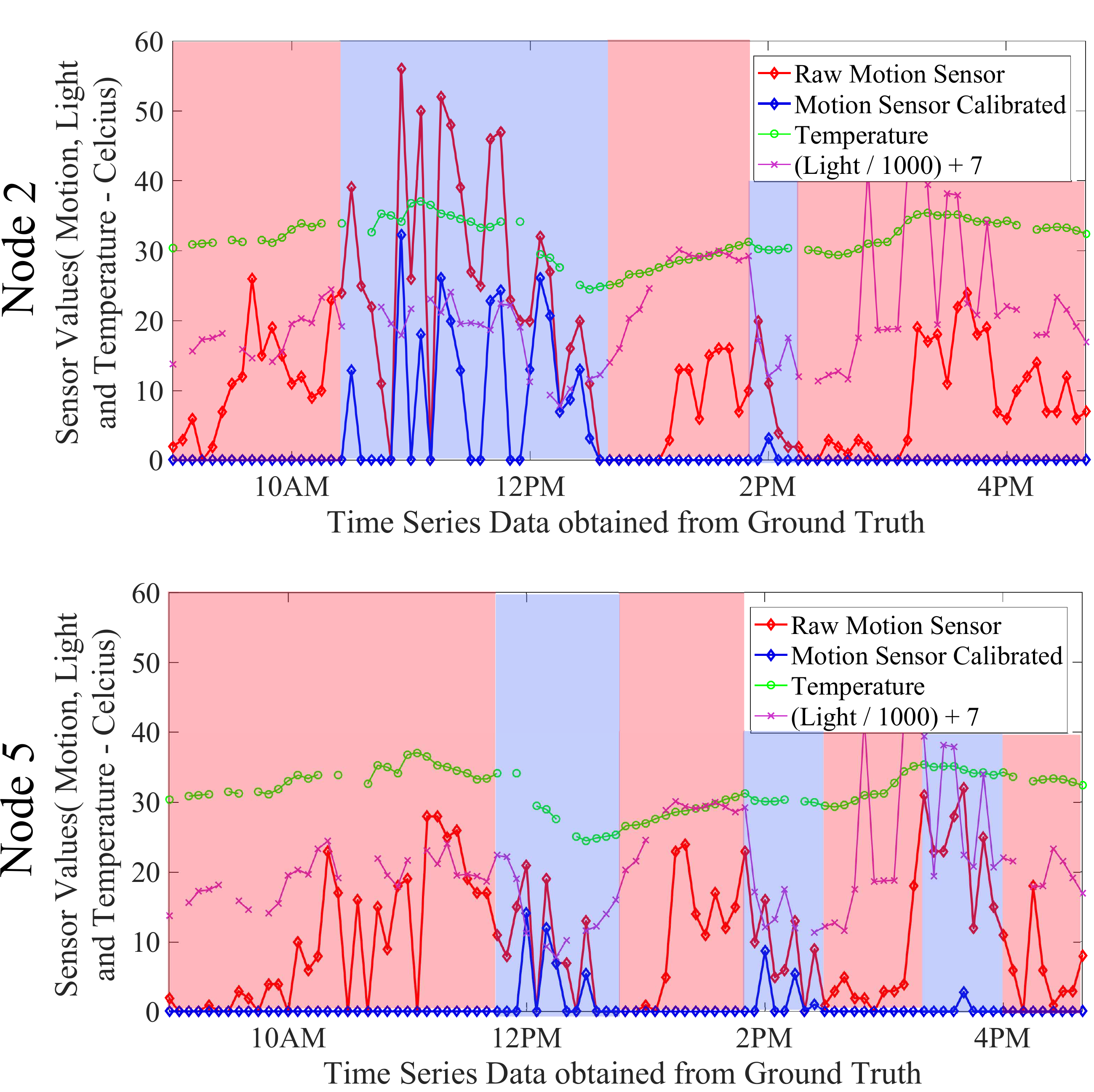}	
	\vspace{-1.0em}	
	\caption{Ground truth obtained on 26 August 2016, where the area highlighted in red indicating no people around and blue color representing people walking through or engaging an activity on that area.}
	\label{fig:CalibratedMotionValue}
\end{figure}
\vspace*{-2mm}

\subsection{Evaluation}
To evaluate the afore-mentioned calibration model, we examine the motion data collected at 12 August 2016 with two different nodes.
Passing by people is recorded manually to check and later is compared with the processed motion data to ensure motion data are calibrated correctly.
The graph in Fig. \ref{fig:CalibratedMotionValue} shows the calibration model compared to the raw motion data and calibrated data along with some highly correlated weather data.
Ground truth is obtained from 10.00AM to 04:00PM and we have highlighted a few zone indicating motion and non-motion value recorded based on different color. 
The raw motion data is plotted in red and it shows tremendous amount of false positive recorded during data collection, where as the blue color indicates a activity is happening within the range of PIR sensor.

In conclusion, utilizing the calibration model proposed earlier, we managed to eliminate the false alarm in the raw motion data while retaining activity detected by PIR sensor. 
\vspace*{-2.0mm}

\subsection{Normalization}
After obtaining processed motion data, normalization is required in order to fuse with sounds sensor. 
Normalization value we chose is based on the distribution of the raw motion data, where only the threshold remain at 85\% of the total data collected. 
Cumulative distribution function (CDF) are used to determine the normalization values.
In this case, we used $\text{normValue} = 10$ as the normalization value for the normalization module. The following Eqn \ref{eqn:normFunction} is the space utilization generated using motion sensor:
\vspace*{-1.0mm}
\begin{equation}
\eta _{{{\cal M}_t}} =\min({\mathcal{M}_{t} \over{\text{normValue}}},1)
\label{eqn:normFunction}
\end{equation}
where $\text{normValue}$ is the normalization value for the motion sensor. 
Since previously, the processed motion data $\mathcal{M}_{t}$ has a lower bound of 0, the value generated by the normalization function will be in the range of $\{0.0,...,1.0\}$.

\vspace*{-2.0mm}

\section{Space Utilization by Sound Sensor}
\label{sec:utilizationSound}

\begin{figure}[ht]
	\centering
	\vspace{-0.5em}
	\includegraphics[width=0.36\textwidth]{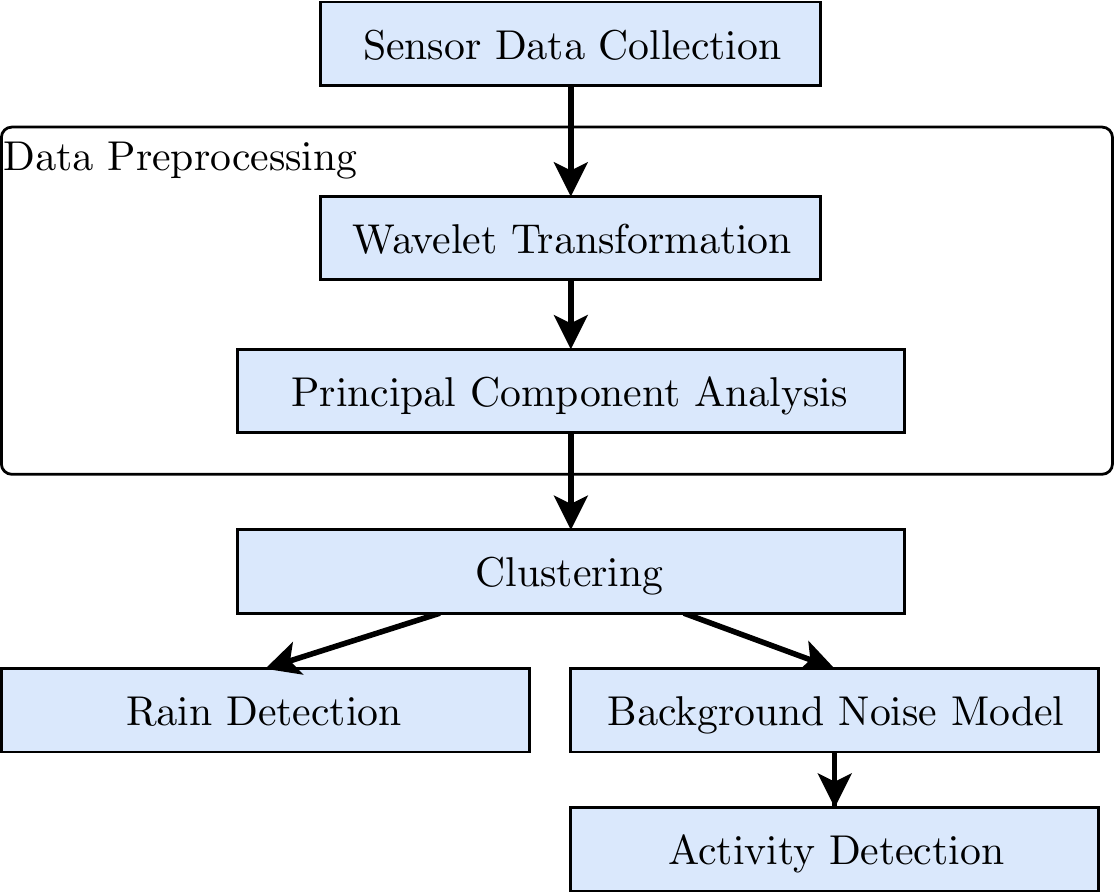}	
	\vspace{-1.0em}
	\caption{Sound Data Processing Model}
	\label{fig:NoiseSystemModel}
\end{figure}

In this section, we will show that how the sound sensor data can be used to estimate the raining period and to detect the occurrence of activities that happened in the vicinity of each sensor nodes. These sound sensors typically capture the noise from sources like people chatting, vehicles, metropolitan trains etc. Summary of our sound sensor data processing model is presented in  Fig. \ref{fig:NoiseSystemModel}. 
As mentioned in the Table \ref{tbl:sensorSpec}, sample sound data is condensed into a histogram, which reduces the overhead incurred during the data acquisition. To this end, each histogram contains the sound information of a five minutes interval.
\vspace*{-2.0mm}

\subsection{Data Prepossessing and Clustering }
Converting the sample data into a meaningful feature is paramount important in a data-mining process. 
In this paper, we use the localized behavior of wavelet transformation for the development of the feature space. \textit{Haar} basis function is used as the mother wavelet due to its own discontinues nature. 
Therefore, we will represent each five minutes histograms in terms of \textit{Haar} basis function.

Due to the high collinearity between histogram bins, principal component analysis (\textit{PCA}) is used to construct a rotated set of orthogonal axes in the variable space. 
This allows us a more robust distance calculation, which is a necessary step for clustering. 
Moreover, by selecting the best $p$ principal components (\textit{PC}), we remove the unwanted noise from the sound data, which possess in \textit{PC}s that captures the relatively low variation of the data. 
In this paper, we select $p$ by calculating the smallest integer value that satisfies.
\vspace*{-1.0mm}
\begin{equation}
\displaystyle \frac{\displaystyle \sum_{i=1}^{m} \| X^{(i)} - \hat X^{(i)} \|^2} {\displaystyle \sum_{i=1}^{m} \| X^{(i)} \|^2}  \leq \alpha, 
\label{eqn:pca}
\end{equation}
\vspace*{-3mm}

where $m$ is the number of samples in the data set, $X^{(i)}$ is the $ i $th element of the dataset, $\hat X^{(i)}$ is the projected value of $X^{(i)}$ using $ p $ principal components, and $ \alpha $ is the given threshold value, and we set $ \alpha $ to 0.95, unless otherwise noted in this paper. 
Therefore, we select the optimal $ p $ number of \textit{PC}s, which minimize the square projection error.

We use clustering as an unsupervised learning method, which allowed us to find the similar patterns of the sound data. In this paper, we consider agglomerative hierarchical clustering, since it can be adaptable for different data sets by using different similarity measurements, \textit{i.e.,} similarity measurement between individual objects and individual clusters. 
In hierarchical clustering, different similarity measurements (between elements and between clusters) shows a different level of distortions in the metric spaces,\footnote{Distances between all sound histograms in each day for a given sensor node.} by the cause of diversity nature\footnote{Since our sound data are significantly varied on different environmental condition on different days.} of the sound data. 
Therefore, it is important to find the correct similarity measurements for the sound data on each day of different sensor nodes, so that we can reduce the distortion of the metric spaces. We calculate the Cophenetic Correlation Coefficient (\textit{CCC}) as a goodness of fit static for the hierarchical dendrogram for each similarity measurement type. 
Then, we select the dendrogram that has the highest \textit{CCC} value, which shows the best clustering consistency. \textit{CCC} can be calculated as
\vspace*{-1.0mm}
\begin{equation}
CCC=\frac{\displaystyle\sum_{i=1}^{n-1}\sum_{j>1}^{n}(c_{ij}-\bar{c})(d_{ij}-\bar{d})}{(\displaystyle\sum_{i=1}^{n-1}\sum_{j>1}^{n}(c_{ij}-\bar{c})^2)^\frac{1}{2}(\sum_{i=1}^{n-1}\sum_{j>1}^{n}(d_{ij}-\bar{d})^2)^\frac{1}{2}},
\end{equation}
where $d_{i,j}$ and $c_{i,j} $ are the individual object distances and the cluster distances respectively.

\begin{figure}[ht]
	\centering
	\vspace{-0.5em}
	\begin{tabular}{c}
		\includegraphics[width=0.46\textwidth]{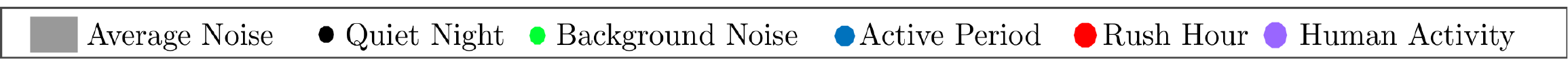}		\\
		{\fontsize{8}{8} \selectfont (a) Typical Weekday }\\
		\includegraphics[width=0.48\textwidth]{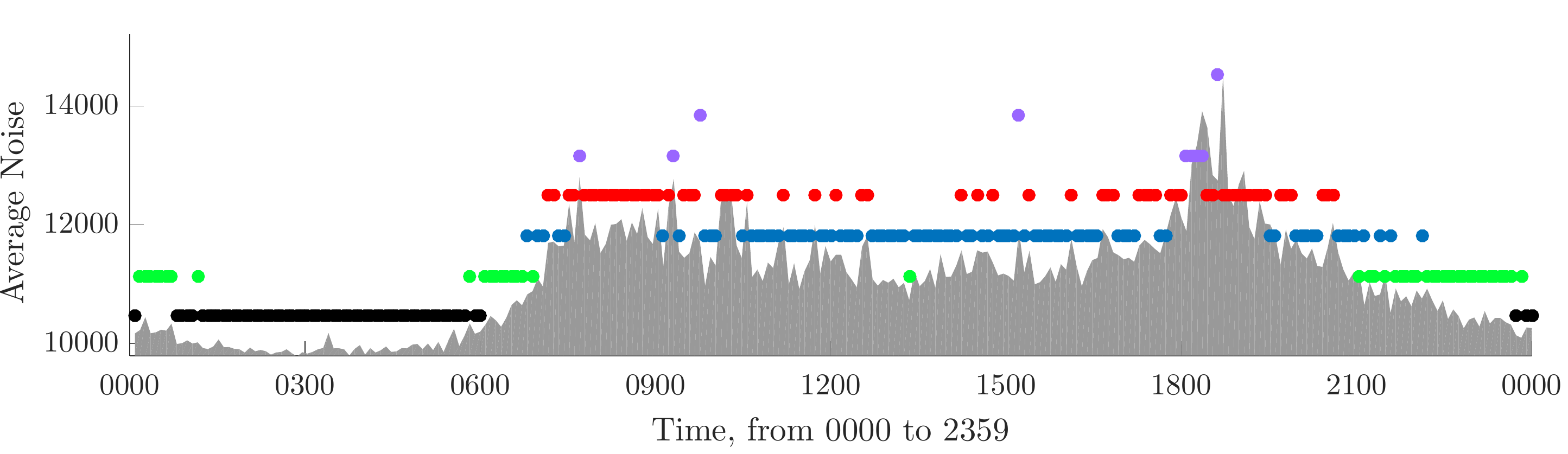}		\\
		{\fontsize{8}{8} \selectfont (b) Typical Saturday} \\
		\includegraphics[width=0.48\textwidth]{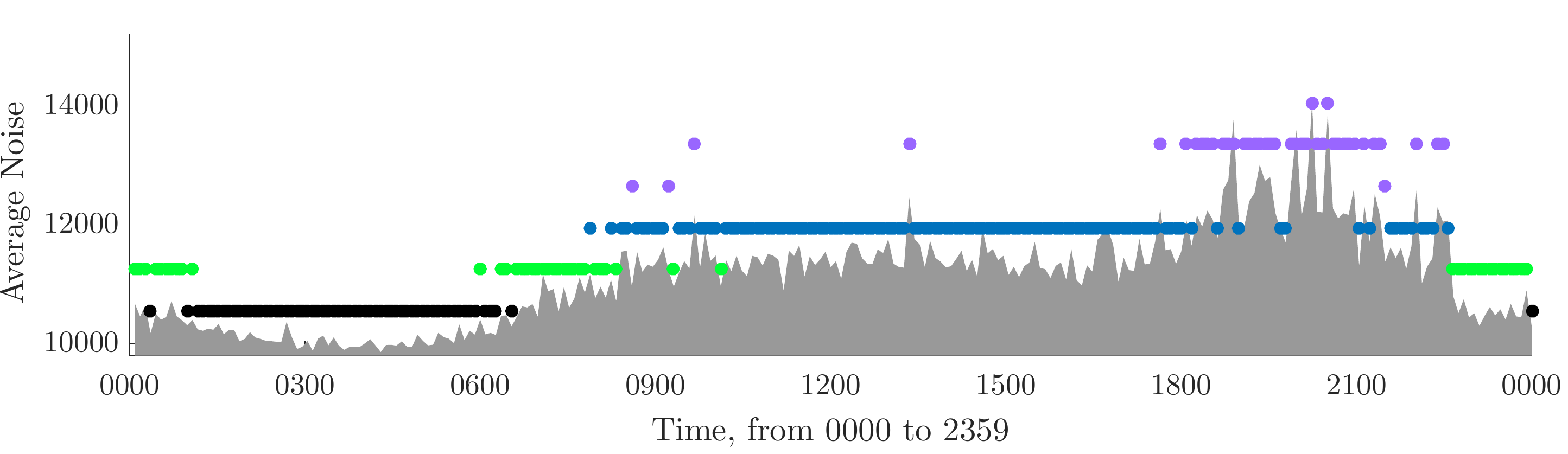}		\\	
		{\fontsize{8}{8} \selectfont (c) Typical Sunday}\\	
		\includegraphics[width=0.48\textwidth]{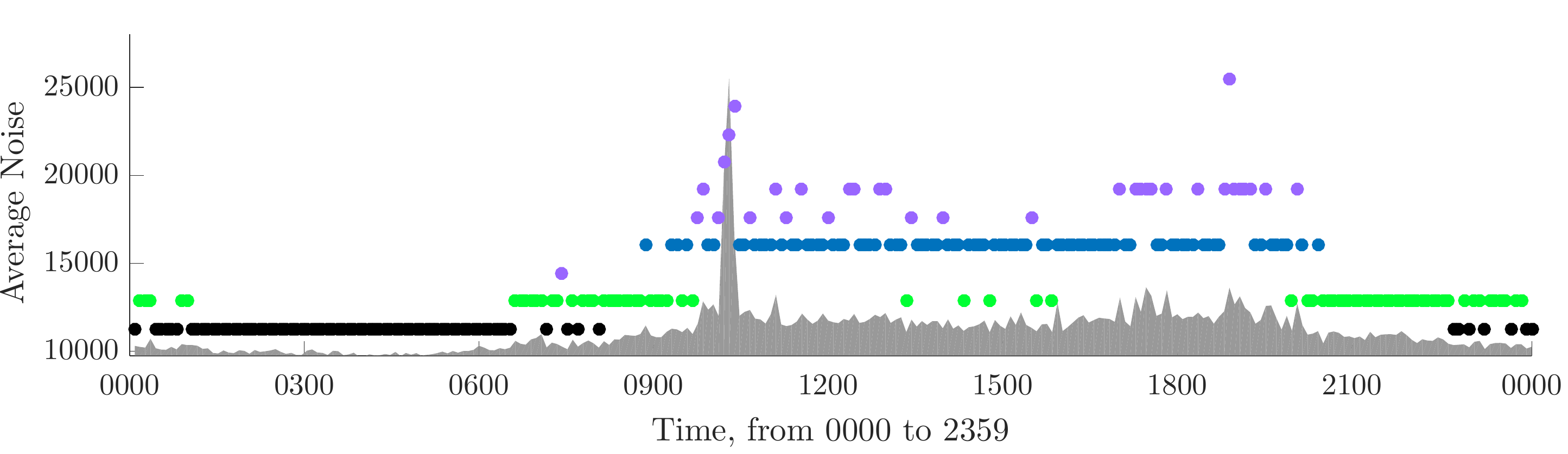}	
	\end{tabular}
	\vspace{-1.0em}
	\caption{Background noise periods detection for each sensor node  in each day separately with comparison of average values. (a) Weekday 2nd August (upper figure). (b) Saturday 16th July (middle figure). (c) Sunday 10th July (bottom figure).}
	\label{fig:backgroundnoise}
\end{figure}

We calculate the optimal number of clusters $N$, which maximizes the \textit{CH} index value. \textit{CH} index is given by
\begin{equation}
CH(N) =  \frac{\displaystyle B(N)/(N-1)}{\displaystyle W(N)/(n-1)},
\end{equation}
where $ B(N) $ is the inter-cluster variation, and $ W(N) $ is the intra-cluster variation for a given $ N $. The intuition behind is that we select the optimal number of cluster $N$, which maximizes the inter-cluster variation while minimizes the intra-cluster variation.

\subsection{Rain Period Detection}

\begin{figure}[htb]
	\centering
	\vspace{-0.5em}	
	\begin{tabular}{c}
		\includegraphics[width=0.4\textwidth]{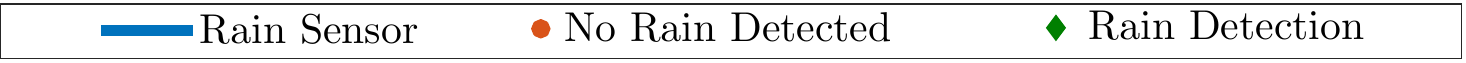}		\\
		\includegraphics[width=0.48\textwidth]{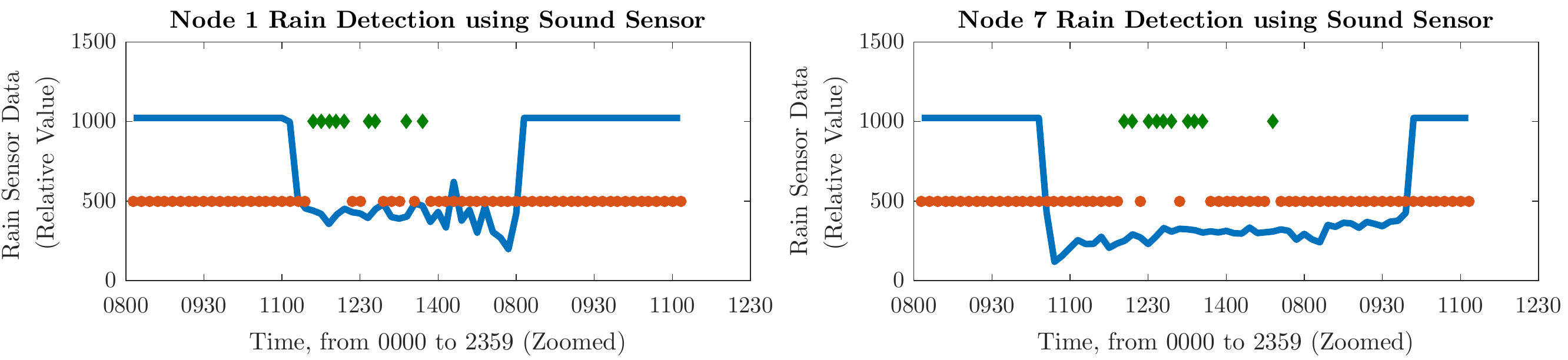}	
	\end{tabular}
	\vspace{-1.0em}
	\caption{Rain detection using noise sensor data and rain sensor data on 28th July 2016, where all 7 Nodes are approximately 5 to 8 meters apart and located within the same playground.}
	\label{fig:rain}
\end{figure}

To detect the raining event, we cluster a number of sensors (Node 1 to Node 7) data, which are all located in a \textit{POI}. As the rain has the impact on all the sensor, such event can be identified when we jointly cluster the data from all the sensor nodes. Fig. \ref{fig:rain} illustrates the experimental results, where the raining duration is represented by filled diamonds. This duration can also be verified from the rain sensor. It should be noted that the rain duration that is detected by the rain sensor is much longer than the actual duration since remaining spilt water after raining is also detected as rain. Moreover, it is not hard to see in Fig. \ref{fig:rain} that rain sensor of Node 7 captures the rain earlier than that of Node 1, where the cause may be due to sheltered of rain at Node 1 by leaves. Using the sound data, which can better identify the actual raining duration from the sound data.

\subsection{Activity Detection}
We define the utilization by sound, whenever a sound sensor captures the activities that happen around the vicinity of the sensor nodes. Since the raw sound data does not provide a fair understanding of activity detection, In this section we will show that how we can exploit the sound sensor data to detect activities.

By clustering each sensor node data in each day separately, we can discover the background noise patterns throughout the day as illustrated in Fig. \ref{fig:backgroundnoise}. Different patterns can be observed on the weekday, Saturday and Sunday.
Therefore we consider each background noise period separately (e.g. active period on a weekday) and calculated as a chi-square value for each histogram so that, we can identify the activities as low probability events. The intuition behind is that histograms, which possess lower probability values represent the activities that deviate significantly from an background noise period. Therefore, by selecting the best $p$ principal components from Eqn. \ref{eqn:pca}, we compute the corresponding chi-square statistic ${\chi}^2_{(i)}$ of $ p $ degrees of freedom as follow:
\vspace*{-1.0mm}
\begin{equation}
{\chi}^2_{(i)} = pc^2_{(1,i)}+pc^2_{(2,i)}+\dots+pc^2_{(P,i)},
\end{equation}
where $pc^2_{(n,i)}$ is the standardized principal component score of the $ i^{th} $ histogram and $n\in {1,...P}$. 
Therefore we identify the histograms that has a lower probability value as activities. \textit{i.e.},
\vspace*{-1.0mm}
\begin{equation}
{\eta _{{{\cal N}_t}}} = \left\{ 
\begin{matrix}
1 &\text{ if } & \chi _{(i)}^2 \ge \beta  \hfill \cr 
0 &\text{ if } & \chi _{(i)}^2 < \beta  \hfill \cr \end{matrix} \right.
\label{eqn:soundsUtz}
\end{equation}
where $\beta$ is an empirical critical value in $ P $ degrees of freedom chi-square distribution and $\eta _{{{\cal N}_t}}$ is the binary value that represents the utilization by sound.

\section{Space Utilization by Sensor Fusion}
\label{sec:combineUtz}

\subsection{Combined Motion and Sounds Utilization}
In this section, we will discuss about the public space utilization using the concept of sensor fusion for both motion and sound sensor.
Motion data uses normalization function shown in the Eqn. (\ref{eqn:normFunction}) in order to generate numbers from $\{0.0,...,1.0\}$ and the noise sensor uses Eqn. (\ref{eqn:soundsUtz}) to generate binary input for activity detection.

To compute the combined space utilization, both motion and sound utilization are fused using the following equation:
\begin{equation}
\eta_{t} = \max(\eta_{\mathcal{M}_{t}}, \eta_{\mathcal{N}_{t}})
\label{eqn:spaceUnorm}
\end{equation}
where $\text{norm}(\mathcal{M})$ is the normalization of motion dataset, and $\eta_{\mathcal{N}_{t}}$ is the sounds utilization computed. The $\max ()$ function will always choose the sensor with the largest value either by motion or sounds sensor and use it as space utilization. Through this, both sensors are able to contribute to public space utilization computation based on their sensor characteristics. 

\subsection{PoI Utilization Evaluation}
To evaluate the space utilization of a specific POI as previously shown in Fig. \ref{fig:MixFigureHardware}(a), we have collected the data over months. 
In particular, we studied data collected starting from 01 to 31 August 2016.
Using the statistical analysis (85\% of the data distribution to determine the normalization and empirical critical value) on the raw data distribution proposed in \cite{lau2016spatial} for motion and sounds sensor, normalized value for motion $\text{norm}(\mathcal{M})$ is set at 10 and the empirical critical value $\beta $ is determined 8.75.

\begin{figure*}[ht]	
	\fontsize{8pt}{8pt}\selectfont
	\centering \includegraphics[trim = 0 3cm 0 0,width=1.0\textwidth]{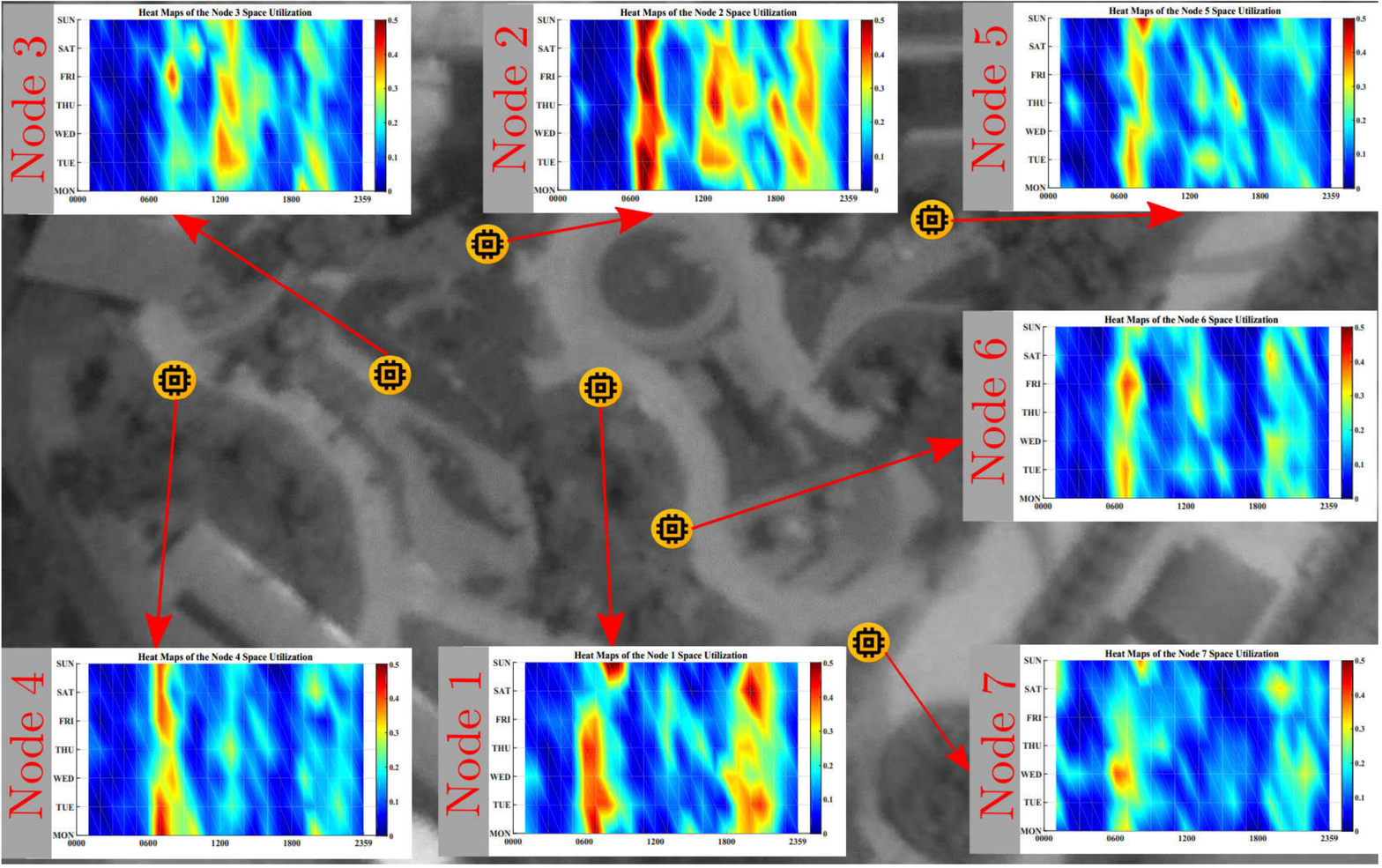}	
	\vspace{-1.0em}
	\caption{Heat maps for all deployed nodes based on combined space utilization for the month of August (01 until 31 August 2016), where the Y axis representing day of the week and X axis representing time in 24 hours. The color scale from blue to red represent the space utilization from low to high (starts from 0.0 to 0.5).}
	\label{fig:rwsnAllUtz}
\end{figure*}

First, we study combined utilization for all the nodes deployed as shown in the Fig. \ref{fig:rwsnAllUtz}.
Compared to other nodes, Node 2 has the highest utilization for the morning and evening session. 
Meanwhile, Node 1 is observed to have more space utilization during night, especially Fri and Weekend. 
Node 1 morning peak (6AM- 7AM) also shifted during weekend, from 8AM in weekday to 9AM in weekend.
As for Nodes 2 and 3, the afternoon utilization seem higher than other nodes in comparison, where Node 3 has less utilization during morning. 
Node 4 has high utilization in the morning, where other sessions' utilization is lower. 
On site investigation reveals that Node 4 is near to a local bus stop, where residents might pass through the Node 4 as a mean of pathway.
Nodes 5, 6, and 7 have similar utilization pattern recorded, but with less utilization as compared with other nodes.

\begin{figure*}[!b]	
	\fontsize{8pt}{8pt}\selectfont
	\begin{tabular}{@{}cccccc}
		\toprule
		& Weekday Average & Weekend Average & Hazy Day & Singapore National Day\\ \midrule
		\multicolumn{1}{c|}{\rot{Motion and Sounds}} 
		&\includegraphics[width=0.22\textwidth]{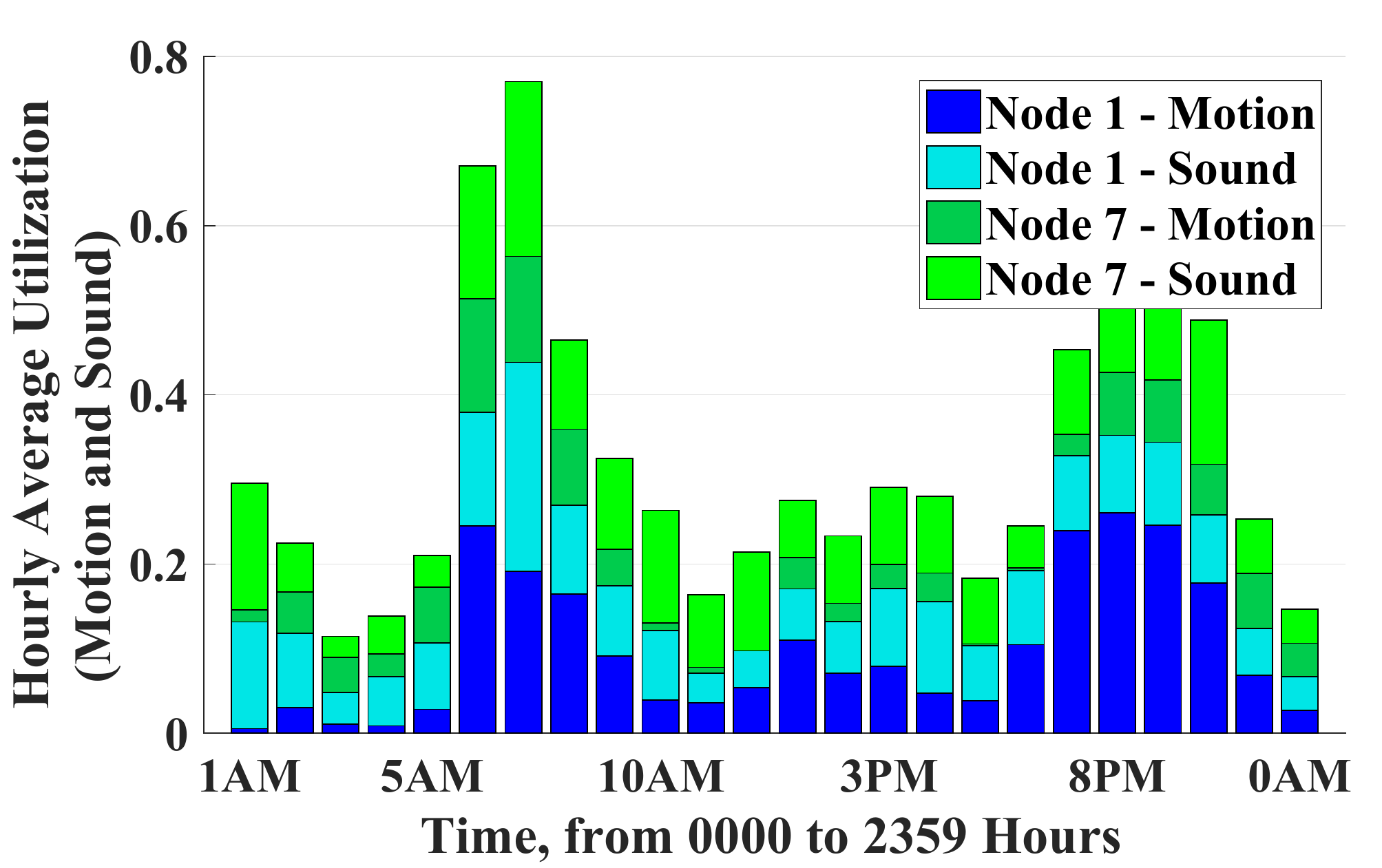}
		&\includegraphics[width=0.22\textwidth]{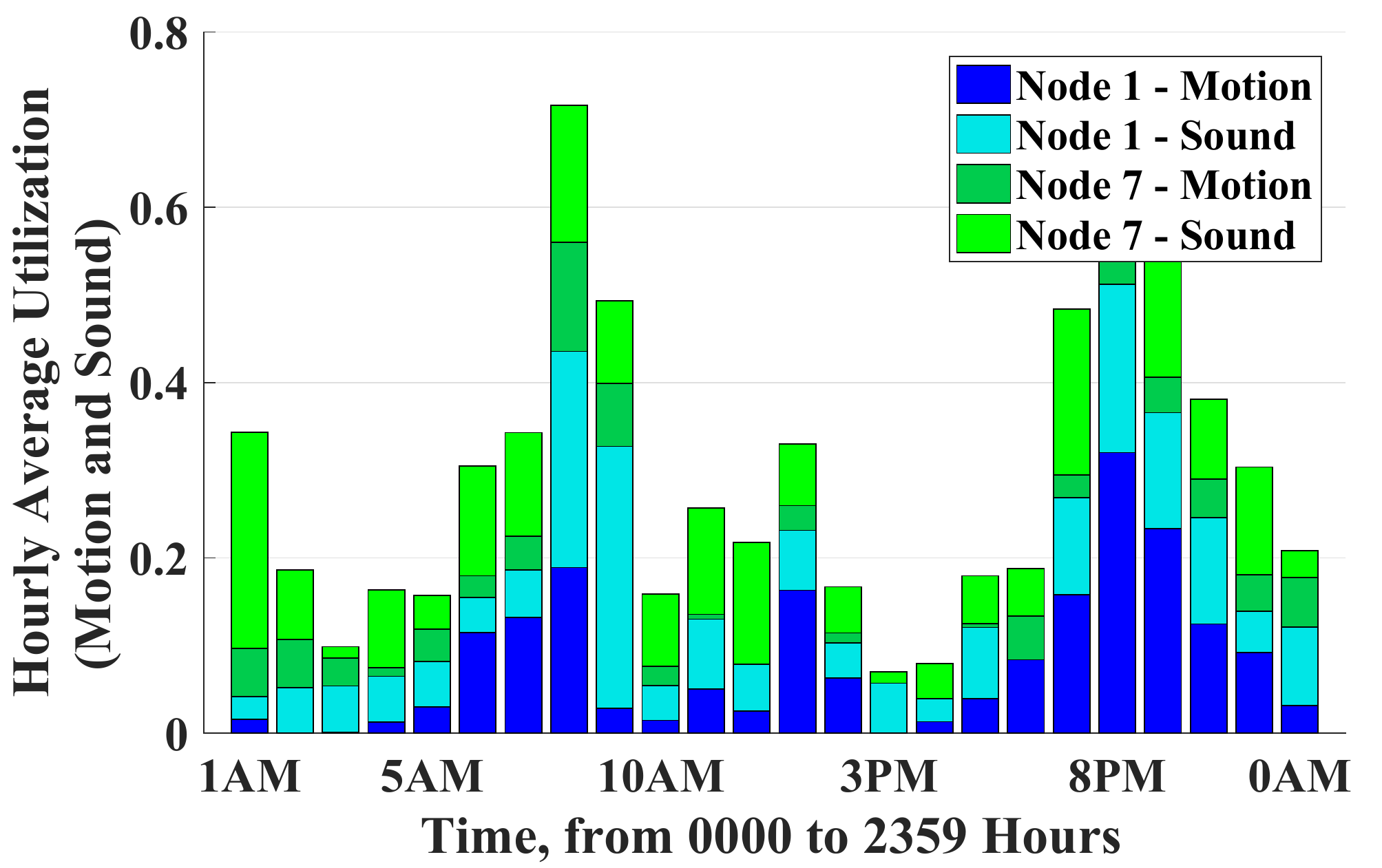}
		&\includegraphics[width=0.22\textwidth]{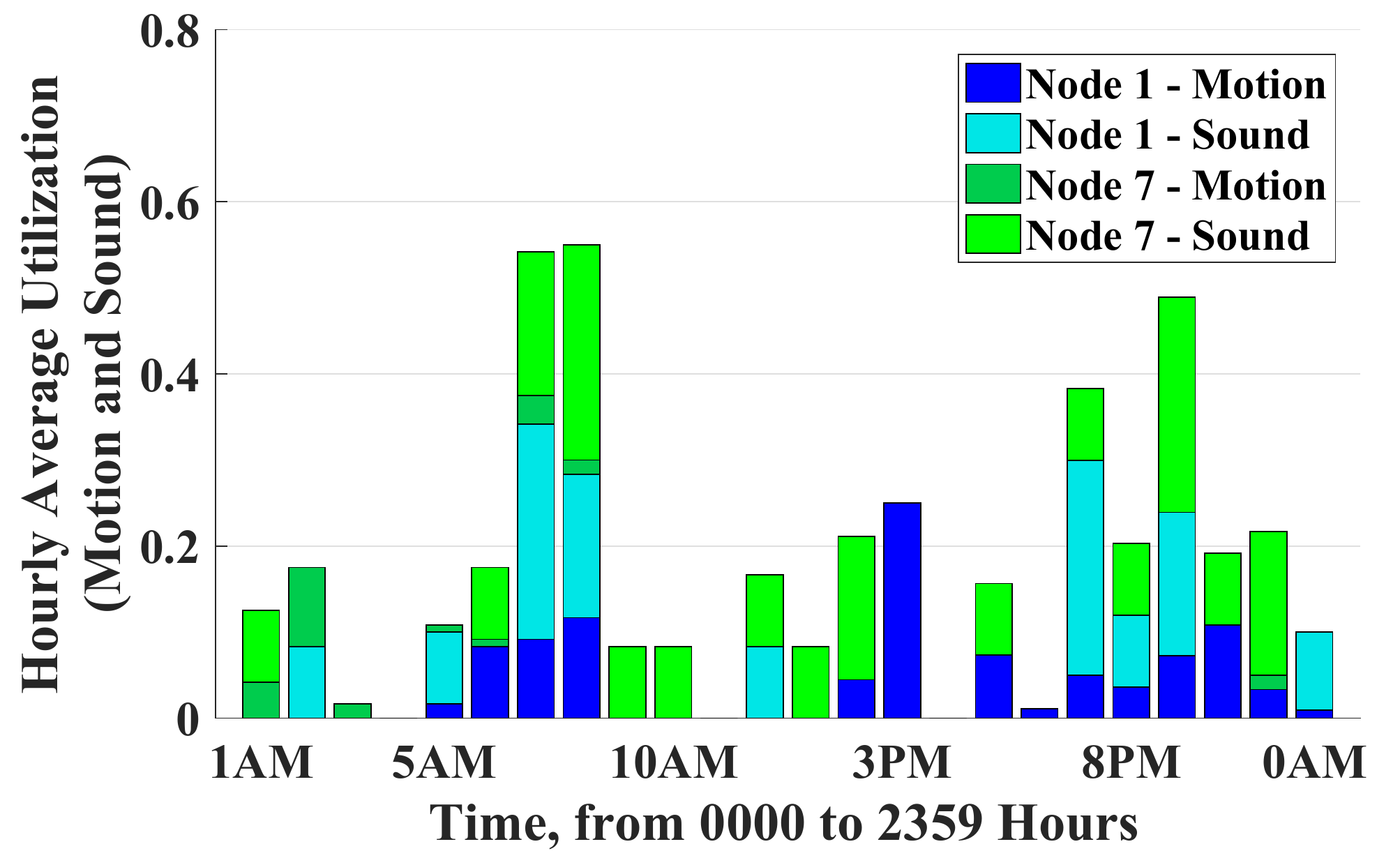}
		&\includegraphics[width=0.19\textwidth]{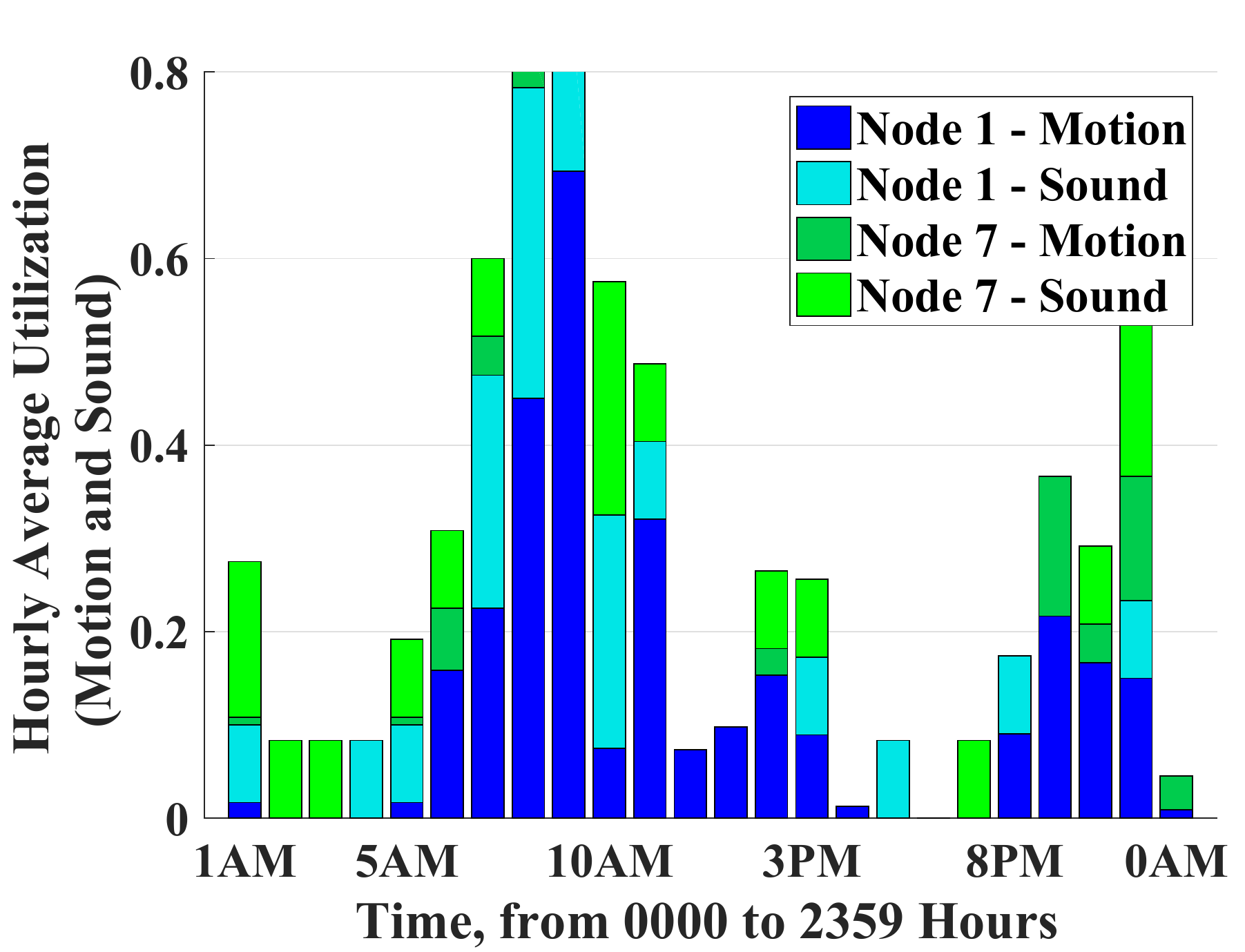} \\
		\multicolumn{1}{c|}{\rot{\hspace{0.8cm}Motion}} 
		&\includegraphics[width=0.22\textwidth]{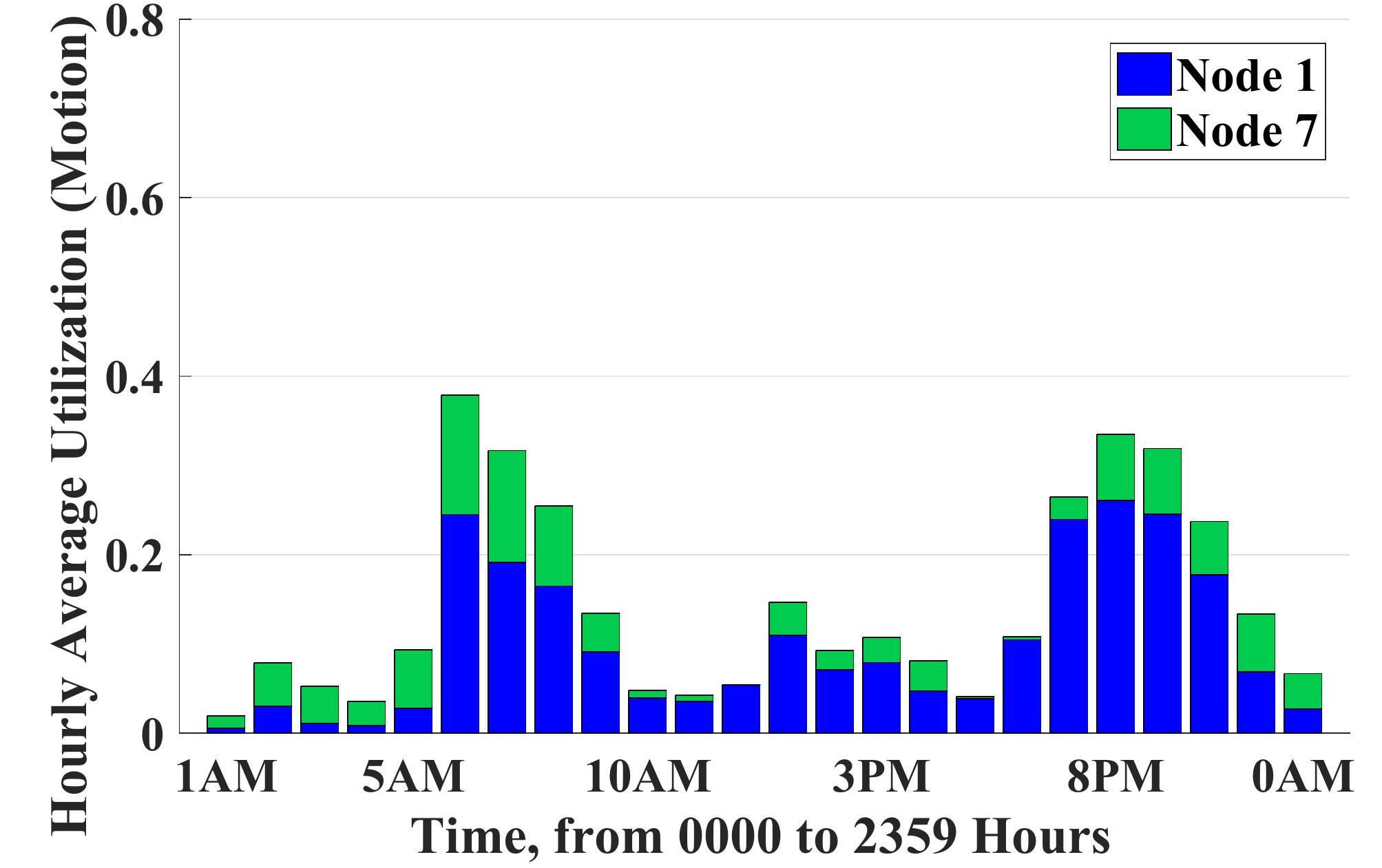}  
		&\includegraphics[width=0.22\textwidth]{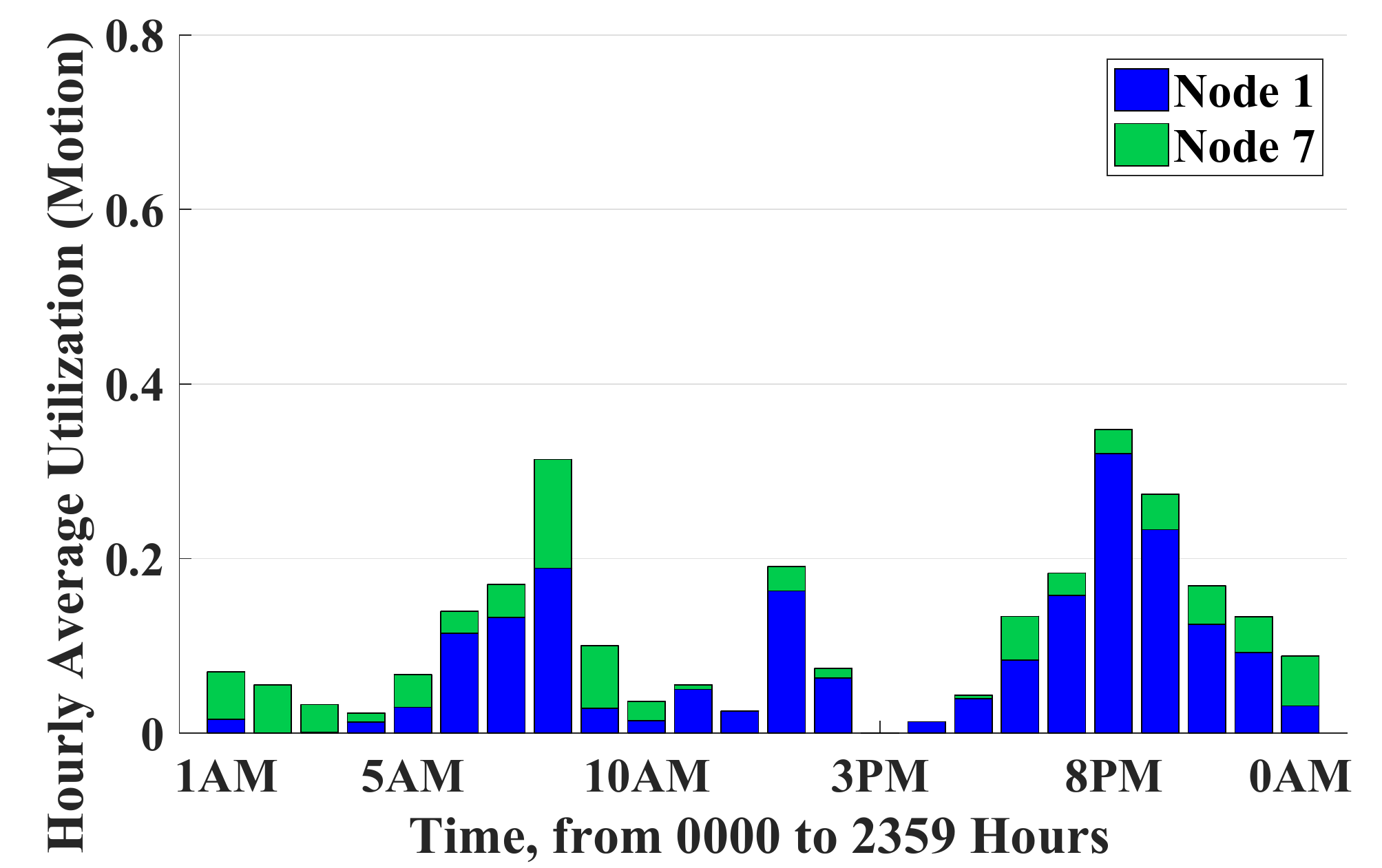}  
		&\includegraphics[width=0.21\textwidth]{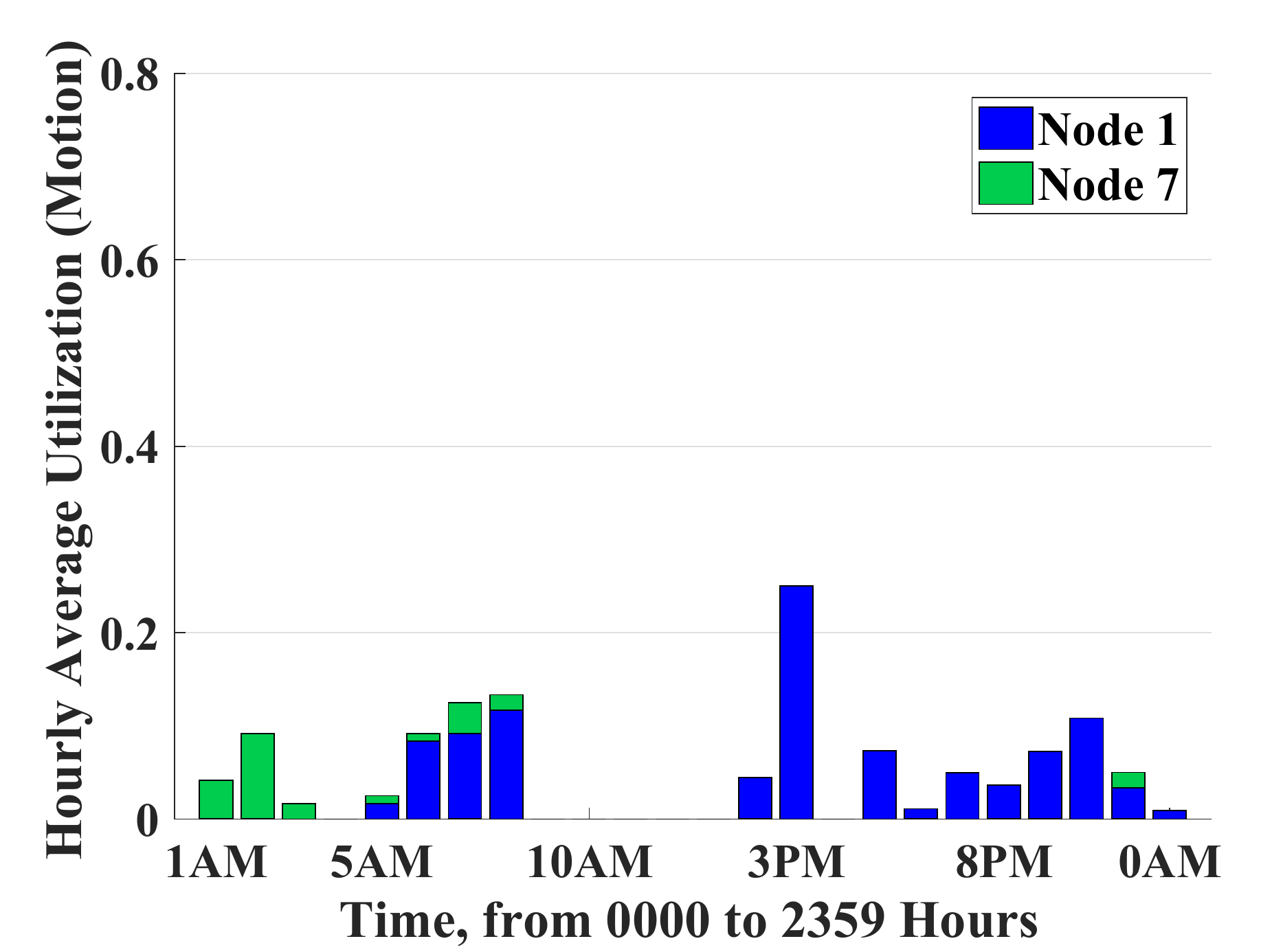}  
		&\includegraphics[width=0.20\textwidth]{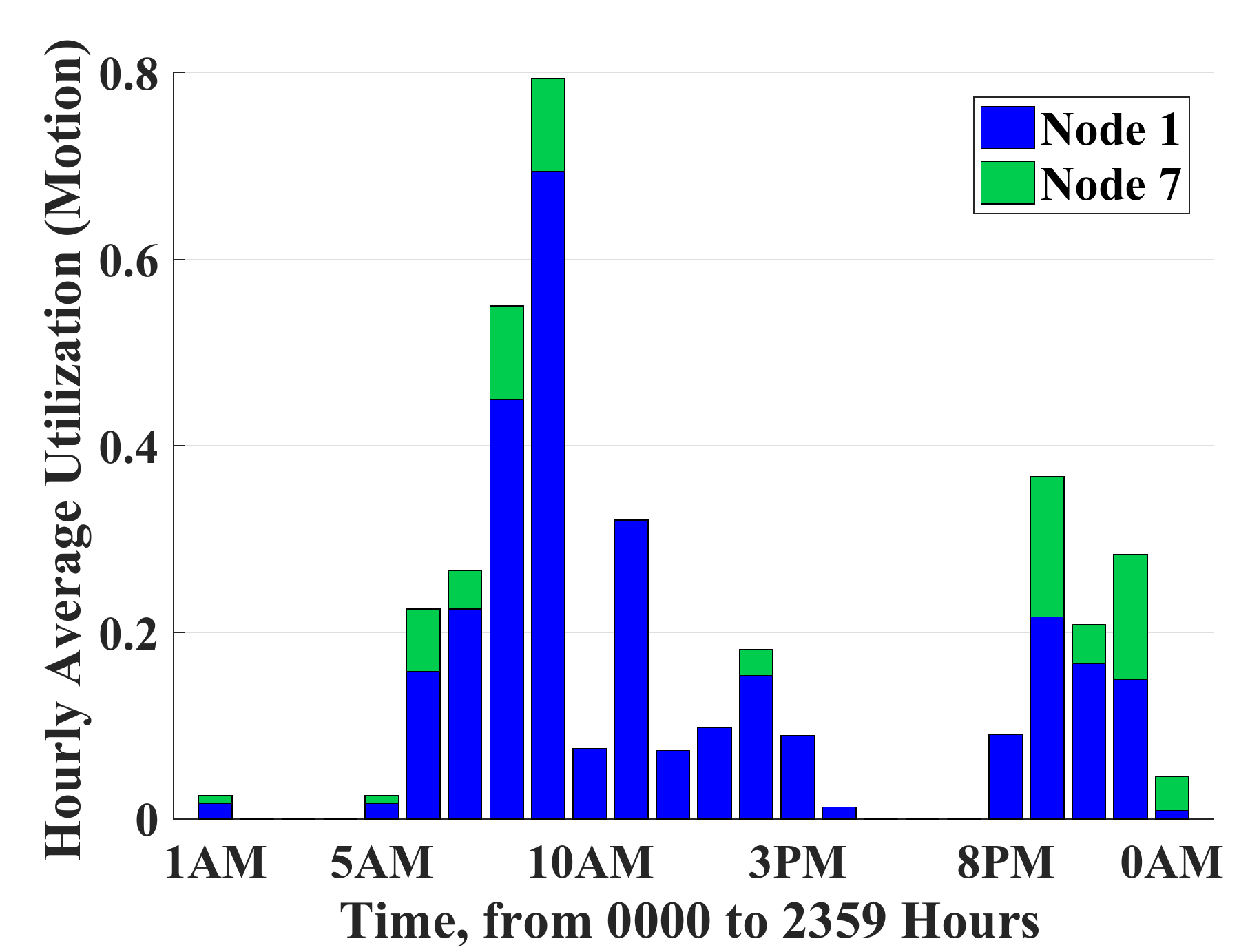} \\
		\multicolumn{1}{c|}{\rot{\hspace{0.9cm}Sounds}} 
		&\includegraphics[width=0.22\textwidth]{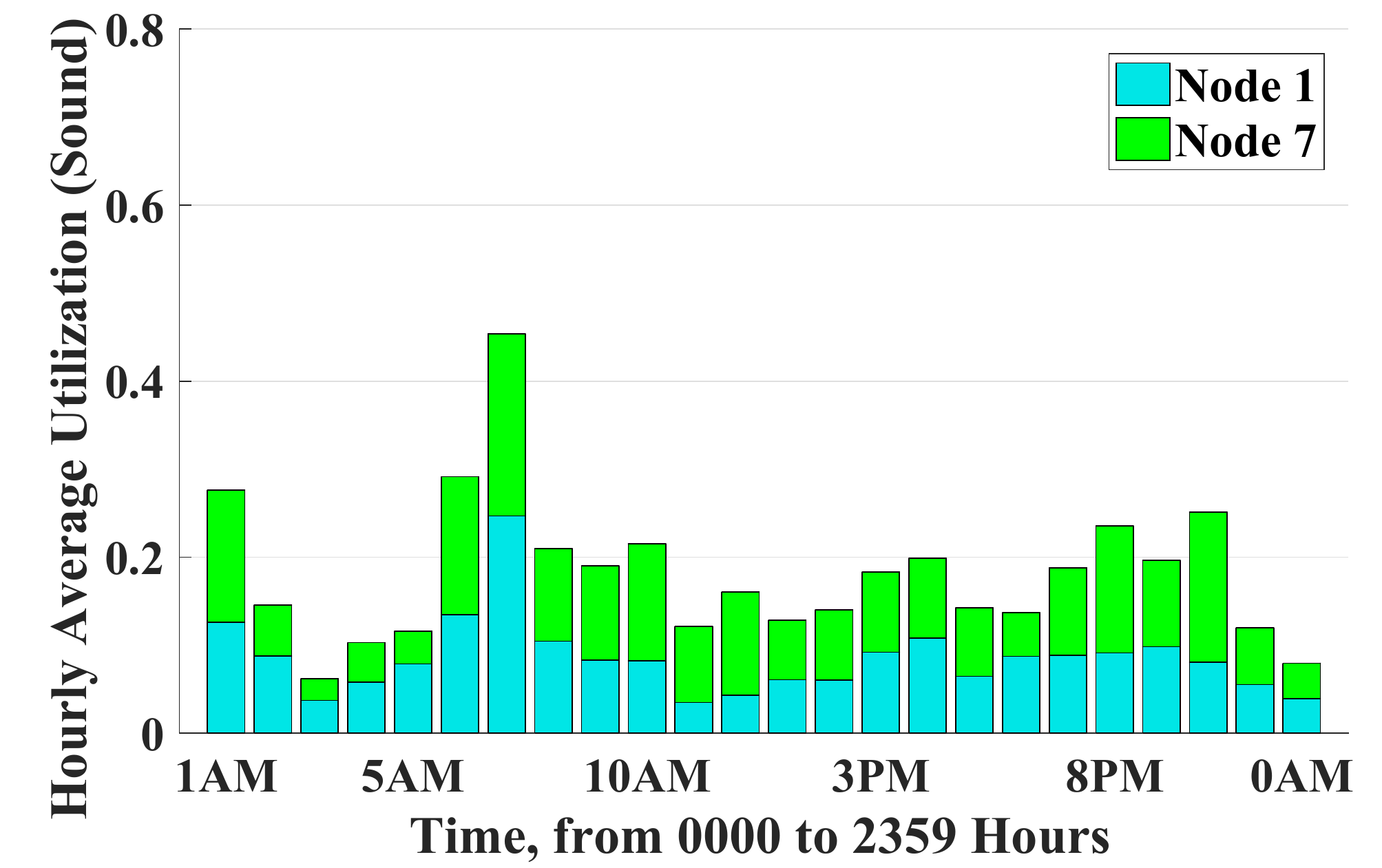} 
		&\includegraphics[width=0.22\textwidth]{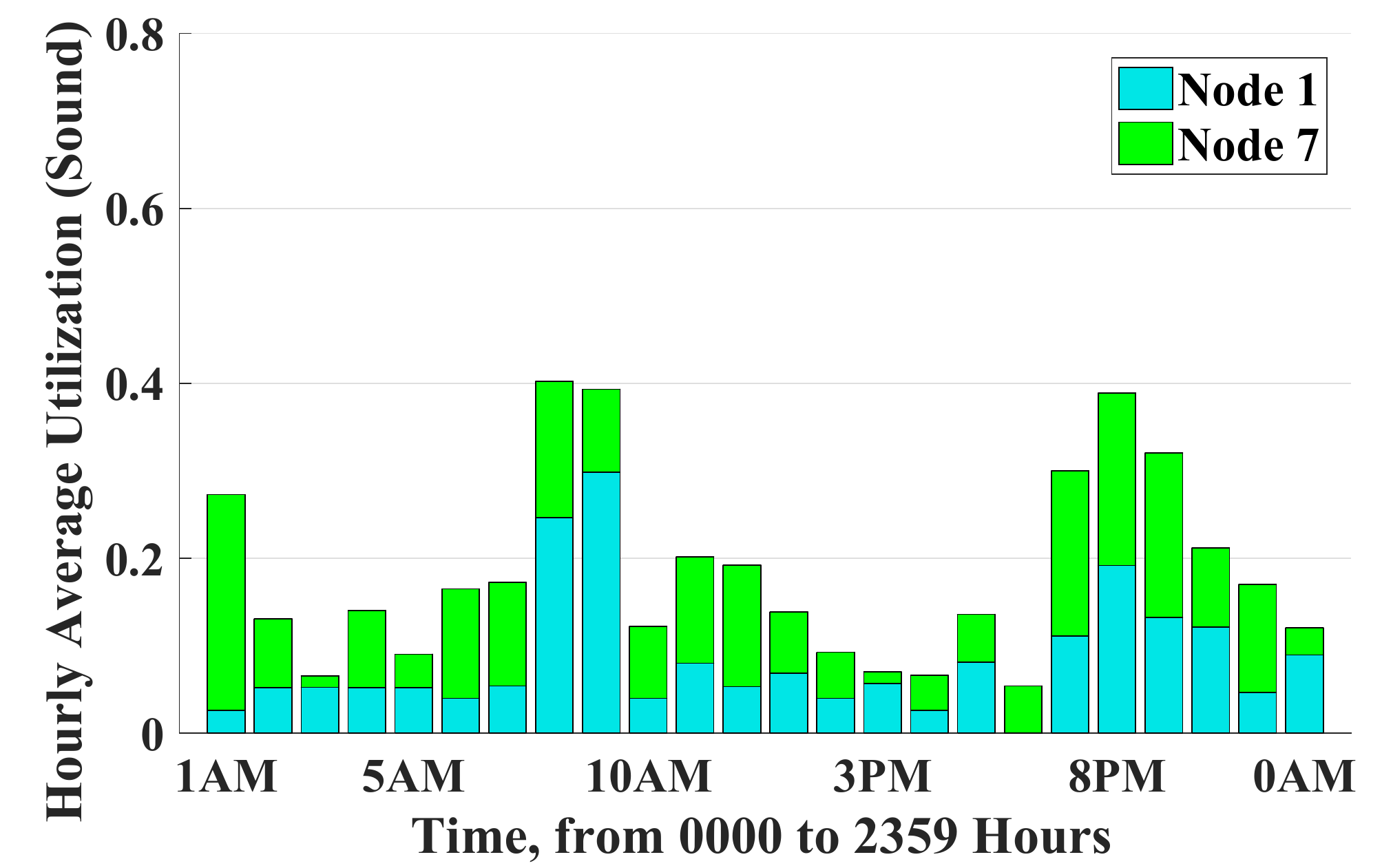} 
		&\includegraphics[width=0.20\textwidth]{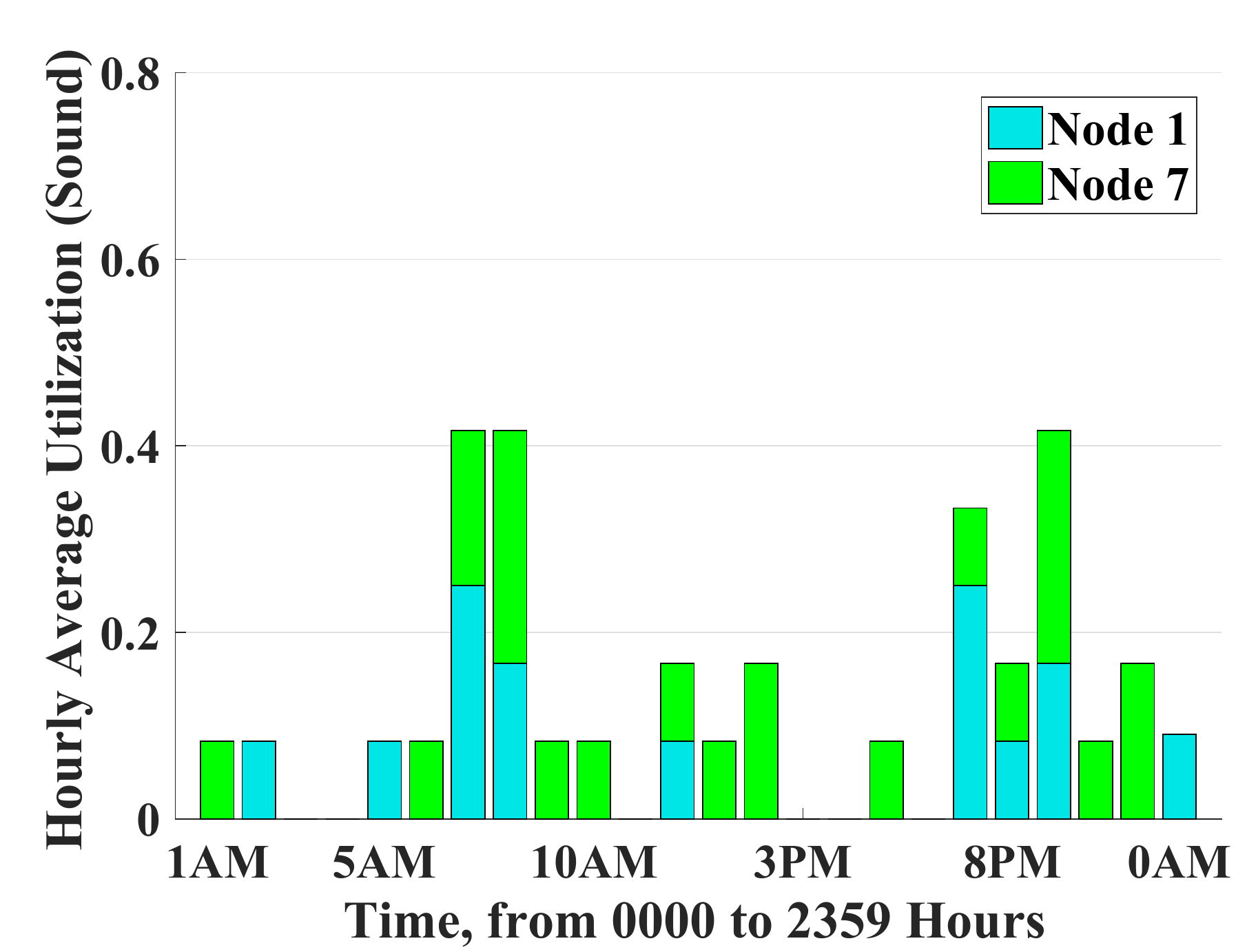} 
		&\includegraphics[width=0.20\textwidth]{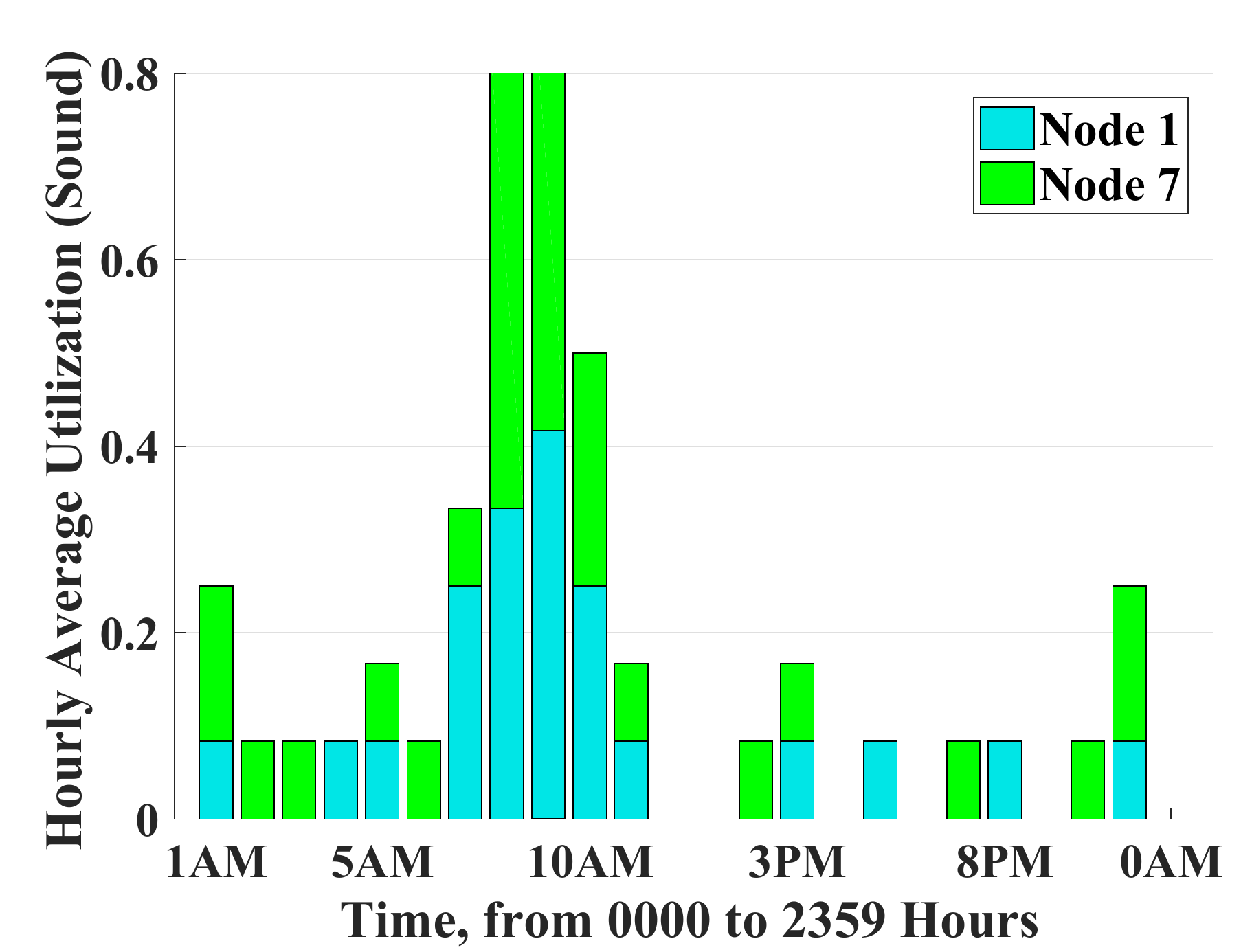}\\ \bottomrule
	\end{tabular}
	\caption{Space Utilization for Node 1 and 7 for 1 month data (from 1st August until 31st August), which are divided into typical average weekday and weekend. Also, two interesting days are studied: Haze Day on 26 August 2016 (around 110PSI on average) and Singapore National Day (09 August 2016)}
	\label{fig:utilizationEvaluation}
\end{figure*}

Based on the previous observation, we can derive that Nodes 1 and 2 have the highest utilization; Nodes 5, 6, and 7 have the lower space usage in overall compared to Nodes 3 and 4. 
To understand the contrast between two utilization pattern, we chose Nodes 1 and 7 to further investigate. 
Using the same datasets, we visualized the data in stacked bar plot as shown in Fig. \ref{fig:utilizationEvaluation}.
In addition, two special days (09 August - Singapore National Day and 26 August - Hazy Day) are also detailed in the figure as well. 

It can be observed that Nodes 1 and 7 share the same utilization pattern for weekday and weekend, which have two peaks in common. 
However, the weekend peak (8AM-9AM) is much later than the weekday peak (6AM-7AM).
From this, we can derive that rush hour for weekend is much later for weekday as residents do not feel the urge of waking up early. 
As we further compared the norm of weekday and weekend against special events such as Singapore National Day and a hazy day, we noticed a few interesting findings. 
First, public space utilization during Singapore National Day is higher than the averaged weekday. 
The morning peak also shift to a later time (8AM-9AM) resembles the weekend utilization pattern, but there is not much evening activity (could be due to the live broadcasting of national day parade ceremony).
Secondly, haze day has a lower utilization compared to any other typical day. 
During haze, pollutant standard index (PSI) reading of 110 on average is recorded for the PoI as shown in \cite{NEA} and it is marked as unhealthy. 
From the deployed Node 1 and 7, space utilization is rather low during haze day. 
It is apparent that residents are reducing outdoor activities, which might end up having less outdoor activity due to hazardous weather condition.

Based on the motion and sound sensor data, both of them produce a similar utilization profile, which can be interpreted as multiple perception of a space utilization. 
Through the fusion of motion and sound utilization is able to produce a more profound model of visualizing usage of a particular POI.



\section{Conclusion}
\label{sec:conclusion}
In this paper, we have proposed a data processing model for RWSN to understand public space utilization. Our model have addressed different challenges such as false positive generated by PIR sensor and characterizing the noise signature from analog sounds sensor. After solving each problems, we then used both motion and sounds sensor in order to capture utilization of a particular space. In our future works, we will report the utilization in a neighborhood region, which includes sensor nodes from multiple different PoIs. The impact of weather and how the utilization across different PoIs change across different period of a year will also be topic of interest.

\section*{Acknowledgment}
This work is supported by Singapore Ministry of National Development (MND) Sustainable Urban Living Program, under the grant no. SUL2013-5, ``Liveable Places: A Building Environment Modeling Approach for Dynamic Place Making” project, and especially appreciate the useful discussion and help from the collaborators from MND, HDB and URA.

\bibliographystyle{IEEEtran}
\newcommand{\BIBdecl}{\setlength{\itemsep}{0.25 em}}
\bibliography{bibSpace}  

\vspace*{-0.36in}
\begin{IEEEbiography}[{\includegraphics[width=1in,height=1.25in,clip,keepaspectratio]{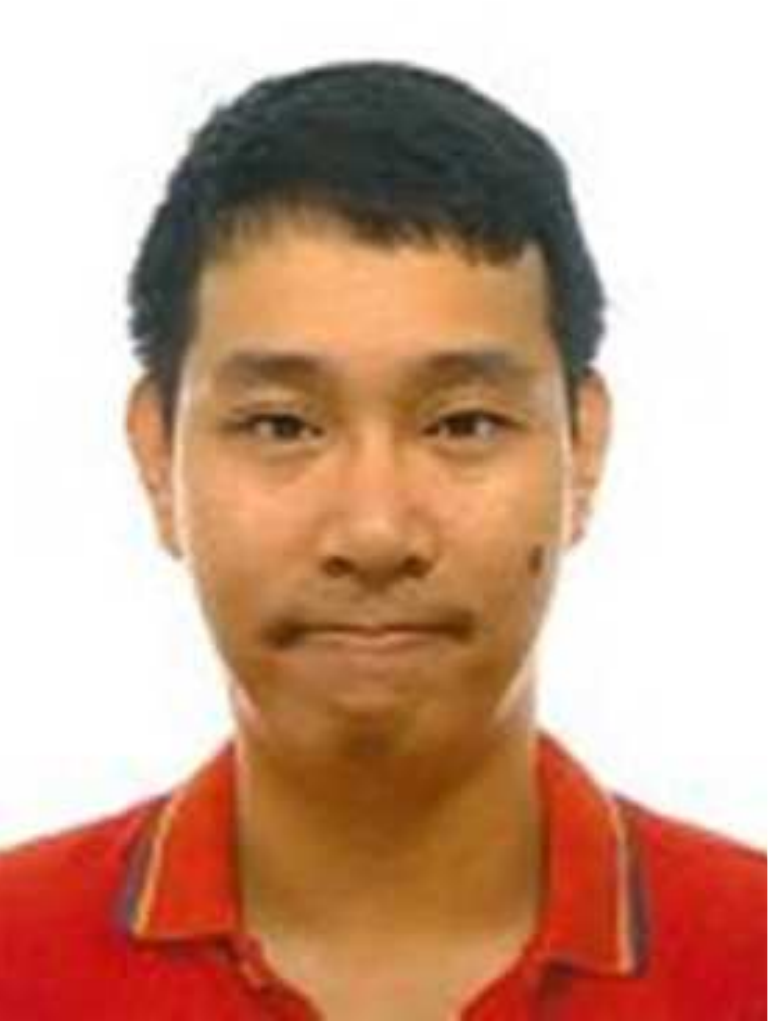}}]{Billy Pik Lik Lau}
received degree in computer science and M. Phil degree in computer science from Curtin University in 2010 and 2014 respectively. He is currently pursuing PhD in Singapore University of Technology and Design (SUTD). He previously works on improving cooperation rate between agents in multi agents systems during master studies and current research focus includes crowd sensing, Internet of things, big data analysis, data discovery, data processing, and unsupervised machine learning. 
\end{IEEEbiography}
\vspace*{-0.6in}
\begin{IEEEbiography}[{\includegraphics[width=1in,height=1.25in,clip,keepaspectratio]{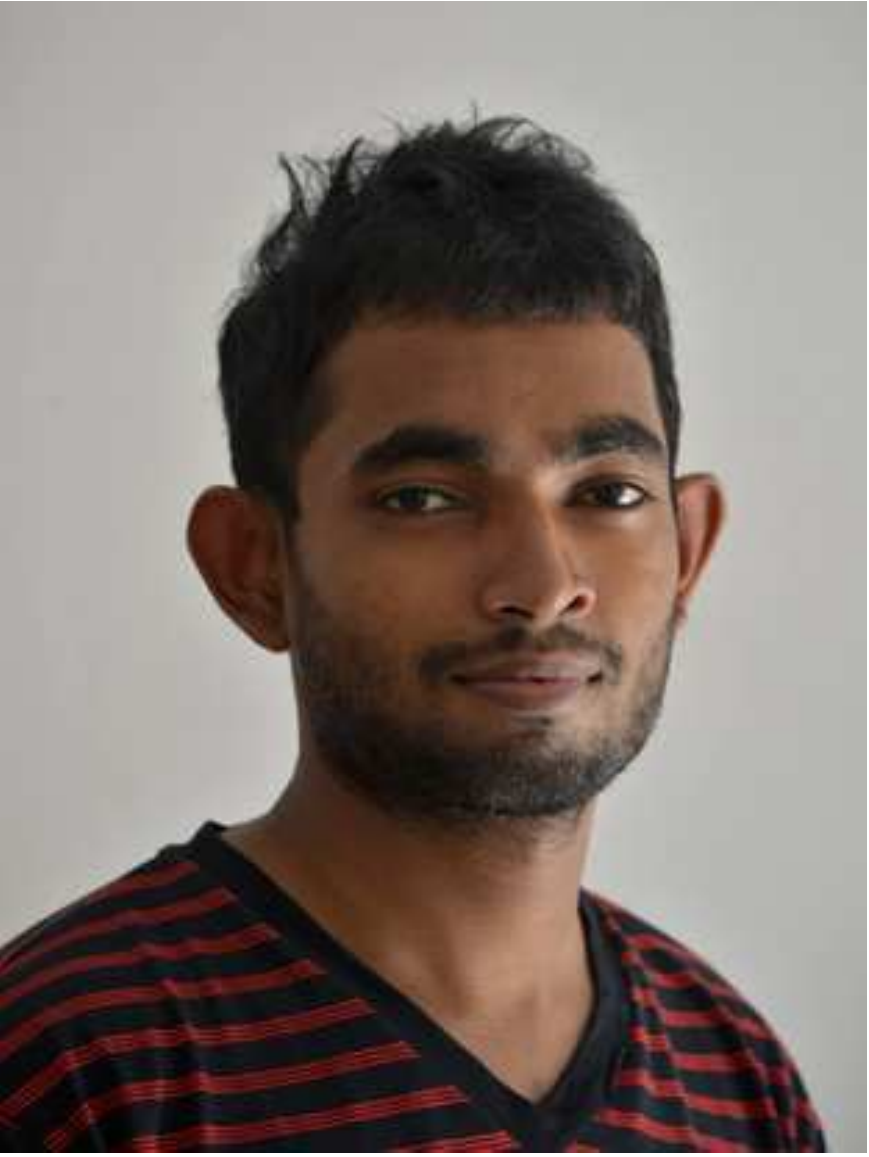}}]{Nipun Wijerathne}
received the B. Sc. degree from the Department of Electronic and Telecommunication Engineering, University of Moratuwa, Sri Lanka in 2016. He then joined with Singapore University of Technology and Design, as a researcher. His research interests include machine learning, deep learning, signal processing, scientific data mining, and pattern analysis methods for practical problems.
\end{IEEEbiography}
\vspace*{-0.5in}
\begin{IEEEbiography}[{\includegraphics[width=1in,height=1.25in,clip,keepaspectratio]{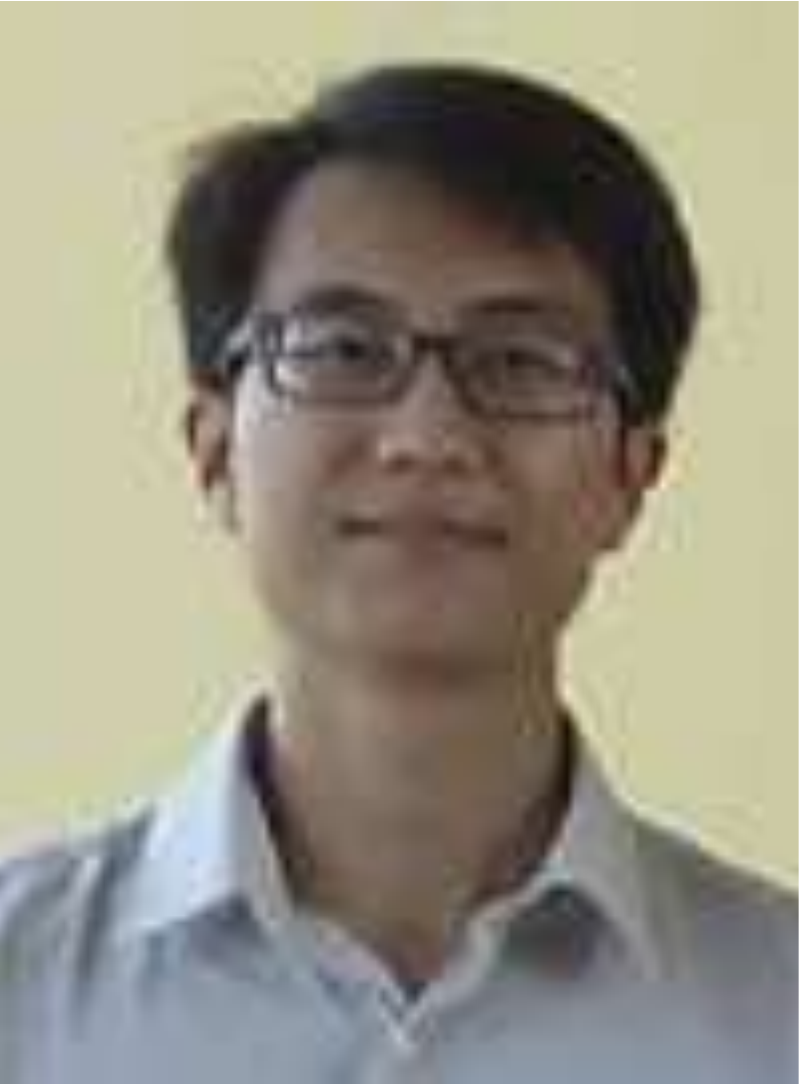}}]{Benny Kai Kiat Ng}
received degree in Electrical and Communication Engineering from Curtin University in 2013. He has been working with Dr Chau Yuen under several research project. His research interest including sensors, data collection, signal processing, hardware, and embedded system development.
\end{IEEEbiography}
\vspace*{-0.5in}
\begin{IEEEbiography}[{\includegraphics[width=1in,height=1.25in,clip,keepaspectratio]{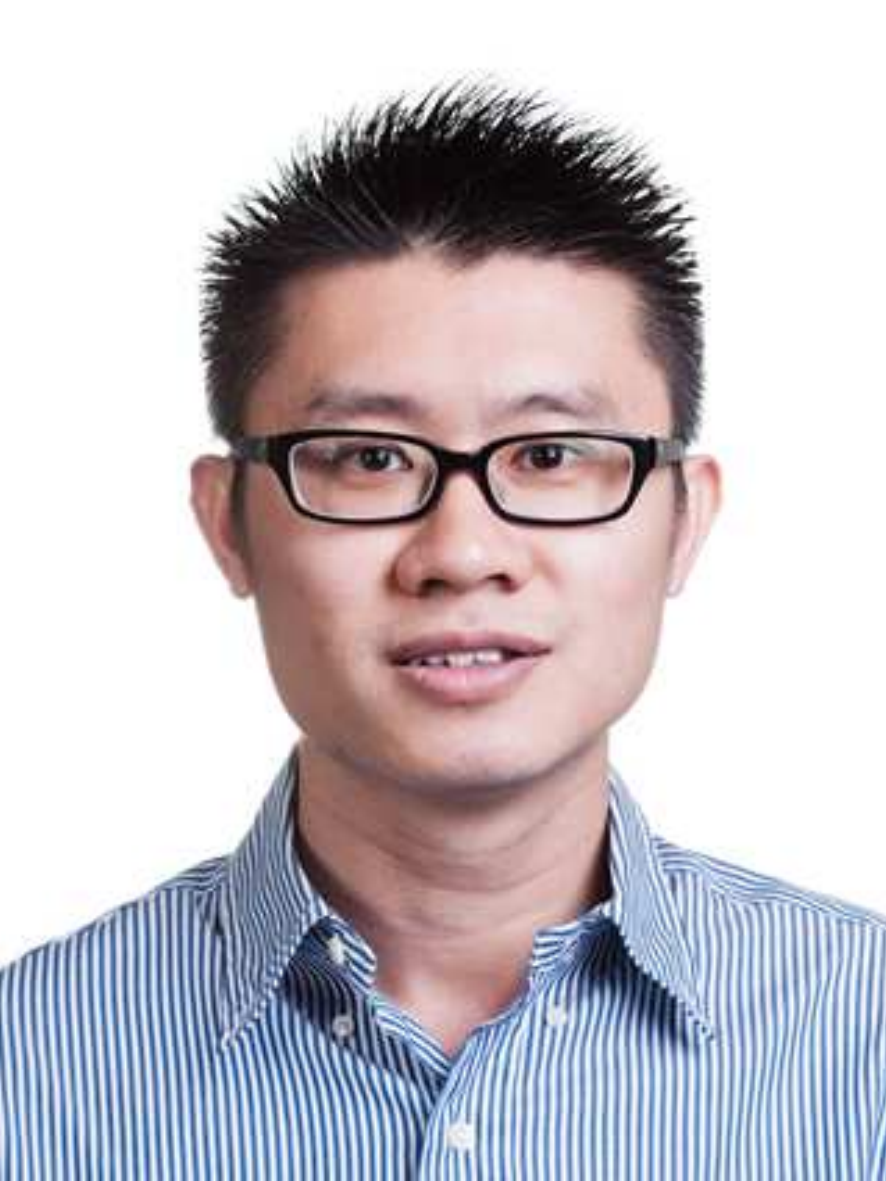}}]{Dr Chau Yuen}
received the BEng and PhD degree from Nanyang Technological University (NTU), Singapore, in 2000 and 2004 respectively. Dr Yuen was a Post Doc Fellow in Lucent Technologies Bell Labs, Murray Hill during 2005. During the period of 2006 ‐ 2010, he worked at the Institute for Infocomm Research (I2R, Singapore) as a Senior Research Engineer, where he was involved in an industrial project on developing an 802.11n Wireless LAN system, and participated actively in o Long Term Evolution (LTE) and LTE‐Advanced (LTE‐A) standardization. He joined the Singapore University of Technology and Design from June 2010, and received IEEE Asia-Pacific Outstanding Young Researcher Award on 2012. Dr Yuen serves as an Editor for IEEE Transactions on Communications and IEEE Transactions on Vehicular Technology. He has 2 US patents and published over 300 research papers at international journals or conferences.
\end{IEEEbiography}


\end{document}